\documentclass[twocolumn,showpacs]{revtex4}

\usepackage{graphicx}
\usepackage{dcolumn}
\usepackage{amsmath}

\begin{document}
\title{Horizon structure of rotating Bardeen black hole and particle acceleration}

\author{Sushant G. Ghosh$^{a,\;b\;}$}
\email{sghosh2@jmi.ac.in}

\author{Muhammed Amir$^{a\;}$}
\email{amirctp12@gmail.com}

\affiliation{$^{a}$Centre for Theoretical Physics,
 Jamia Millia Islamia,  New Delhi 110025
 India}
\affiliation{$^{b}$Astrophysics and Cosmology Research Unit,
 School of Mathematics, Statistics and Computer Science,
 University of KwaZulu-Natal, Private Bag X54001,
 Durban 4000, South Africa}

\begin{abstract}
We investigate the horizon structure and ergosphere in a rotating Bardeen regular black hole, which has an additional parameter ($g$) due to magnetic charge, apart from mass ($M$) and rotation parameter ($a$). Interestingly, for each value of parameter $g$, there exist a critical rotation parameter ($a=a_{E}$), which corresponds to an extremal black hole with degenerate horizons, while for $a<a_{E}$ describes a non-extremal black hole with two horizons, and no black hole for $a>a_{E}$. We find that the extremal value $a_E$ is also influenced by the parameter $g$ and so is the ergosphere. While the value of $a_E$ remarkably decreases when compared with the Kerr black hole, the ergosphere becomes more thick with increase in $g$. We also study collision of two equal mass particle near the horizon of this black hole, and explicitly bring out the effect of parameter $g$. The  center-of-mass energy ($E_{CM}$) not only depends on rotation parameter $a$, but also on parameter $g$. It is demonstrated that the $E_{CM}$ could be arbitrary high in the extremal cases when one of the colliding particle has critical angular momentum, thereby suggesting that the rotating Bardeen regular black hole can act as a particle accelerator.
\end{abstract}

\maketitle

\section{Introduction}
The spherically symmetric Reissner-Nordstr$\ddot{o}$m metric \cite{Reissner:1916} is given by
\begin{equation}
ds^2=g_{\mu \nu} \otimes dx^{\mu} \otimes dx^{\nu},\;\;\;\;(\mu,\nu=0,1,2,3),
\end{equation}
with $g_{\mu \nu}= \text{diag} (-f(r),f(r)^{-1},r^2,r^2\sin^2 \theta)$,
and
\begin{equation}
f(r)=1-\frac{2m}{r}+\frac{q^2}{r^2},\nonumber
\end{equation}
where $m$ and $q$ respectively denotes mass and charge. The solution represent a black hole shielding a singularity, the black hole is being formed as end state of collapse of a star. The gravitational collapse of a sufficiently massive star ($\sim3.5M_{\odot}$) inevitably leads to a spacetime singularity is a fact established by elegant theorem due to Hawking and Penrose \cite{Hawking:1970}, (see also Hawking and Ellis \cite{Hawking:1973}).

However, it is widely believe that these singularities do not exist in Nature, but they are creation or artifact of classical general relativity. The existence of singularity by its very definition, mean spacetime fails to exists signaling a breakdown of physics laws. Thus, in order to exist these laws, singularities must be substituted by some other objects  in more suitable theory. The extreme condition, in any form, that may exist, at the singularity, imply that one should rely on quantum gravity, which are expanded to resolve these singularity \cite{Wheeler:1964}. While we do not yet have any definite quantum gravity, to understand inside of the black hole and resolve it separately, hence we must turn our attention to regular models, which are motivated by quantum arguments. The earliest idea, in mid sixties, due to Sakharov \cite{Sakharov:1966} and Gliner \cite{Gliner:1966} suggest that singularities could be avoided by matter, i.e., with a de Sitter core, with equation of state $p=-\rho$. This equation of state is obeyed by cosmological vacuum and hence, $T_{\mu \nu}$ takes a false vacuum of the form $T_{\mu \nu}=\Lambda g_{\mu \nu}$, $\Lambda$ is cosmological constant.

Thus spacetime filled with vacuum could provide a proper discrimination at final stage of gravitational collapse, replacing future singularity \cite{Gliner:1966}. The first regular black holes solution, based on this idea, was proposed by the Bardeen \cite{Bardeen:1968}, in which there are horizons but no singularity. The matter field is kind of magnetic field. The solution yielding a modification of the Reisnner-Nordstr$\ddot{o}$m black hole solution, with the metric function $f(r)$ being
\begin{eqnarray}
f(r)&=&1-\frac{2mr^2}{(r^2+g^2)^{3/2}}\nonumber \\
&=&1-\left(\frac{m}{g}\right)\frac{2(r/g)^2}{(1+(r/g)^2)^{3/2}},\;\;\;\;
\text{and}\;\;\;\; r\geq0.\nonumber
\end{eqnarray}
Numerical analysis of $f(r)=0$ reveals a critical value  $\chi^{*}$ such that $f(r)$ has  double root if $\chi=\chi^{*}$, two roots if $\chi<\chi^{*}$ and no root if $\chi>\chi^{*}$, with $\chi=m/g$. These cases illustrate, respectively, an extreme black hole with degenerate horizons, a black hole with Cauchy and event horizons, and no black hole.

The Bardeen solution is regular everywhere which can be realized from the scalar invariants
\begin{eqnarray}
R_{ab}R^{ab}&=&\frac{6mg^2(4g^2-r^2)}{(r^2+g^2)^{7/2}},\nonumber \\
R_{abcd}R^{abcd}&=&\frac{12m^2}{(r^2+g^2)^7}\Big[8g^8-4g^6r^2+47g^4r^4\nonumber \\ &-&12g^2r^6+48r^8\Big],
\end{eqnarray}
which are well behaved everywhere including at $r=0$. However, for Reissner-Nordstr$\ddot{o}$m case ($g=0$), they diverges at $r=0$ indicating scalar polynomial singularity \cite{Hawking:1973}. The Bardeen black hole is asymptotically flat, and near origin it behaves as the de Sitter, since \[f(r)\approx 1 - \frac{2m}{g}r^2,\;\; r\approx 0^{+}\] and whereas for large  $r$,  it serve as Schwarzchild.

Thus the black hole interior does not result into a singularity but develops a de Sitter like region, eventually settling with regular centre, thus it's maximal extension of Reissner-Nordstr$\ddot{o}$m spacetime but with a regular centre \cite{Borde:1994ai,Borde:1996df}. There has been an enormous advancement on the analysis, and application of regular black holes \cite{AyonBeato:1998ub,Hayward:2005gi,Bronnikov:2000vy,Zaslavskii:2009kp,Lemos:2011dq}, however all subsequent regular black holes are based on Bardeen idea. The detailed study of circular geodesics of photons of non-rotating regular black hole can be found in \cite{Stuchlik:2014qja}. The deflection of light ray and gravitational lensing in regular Bardeen spacetime is also studied \cite{Schee:2015nua}. The ghost images of Keplerian discs, generated by the photons with low impact parameters for regular Bardeen black hole is also discussed \cite{Schee:2015nua}.

The no-hair theorem suggest that astrophysical black hole candidates are Kerr black hole, it still lacks direct evidence and actual nature is not yet verified.  This open an arena for investigating properties for black holes that differ from Kerr black holes. Lately, the rotating (spinning) counterpart of Bardeen's black hole is proposed, which can be written in Kerr-like form in Boyer-Lindquist coordinates \cite{Bambi:2013ufa}. The rotating Bardeen regular metric has been tested with black hole candidate in Cygnus X-1 \cite{Bambi:2014nta}, and thereafter more rotating regular black holes were proposed \cite{Ghosh:2014pba,Toshmatov:2014nya,Amir:2015pja,Ghosh:2014hea}. Interestingly the 3$\sigma$-bound $a_{*}>0.95$ \cite{Gou:2011} and $a_{*}>0.983$ \cite{Gou:2013dna} for the Kerr black hole, changes respectively to, $a_{*}>0.78$ and $\mid \chi \mid<0.41$, and $a_{*}>0.89$ and $\mid \chi \mid<0.28$ for the rotating Bardeen regular black hole.   Further, the measurement of the Kerr spin parameter of rotating Bardeen regular black holes from the shape of the shadow of a black holes was also explored \cite{Li:2013jra}. The rotating Bardeen black holes, accommodates the  Kerr black hoes as the special case when the deviation parameter, $g =0$, may be regarded as  a well-suited framework for  exploring astrophysical black holes.

In this paper, we investigate the horizon structure and ergosphere of the rotating Bardeen regular black hole and explicitly bring out the effect of the Bardeen's charge $g$.   The paper is organised as. In Sec.~\ref{bh}, we review the rotating Bardeen regular black hole and analyze  horizon structure and ergosphere, with respect to charge $g$. We analyse the equatorial equations of motion of particles and effective potential in Sec.~\ref{eqms}. The Sec.~\ref{cme} is devoted to the collision of two equal masses particles in the background of rotating Bardeen regular black hole, and numerically calculate $E_{CM}$ in near horizon particle  collision, and finally summarize our results and evoke some perspectives to end the paper in Sec.~\ref{conclusion}.

\begin{figure*}
	\begin{tabular}{c c}
		\includegraphics[scale=0.62]{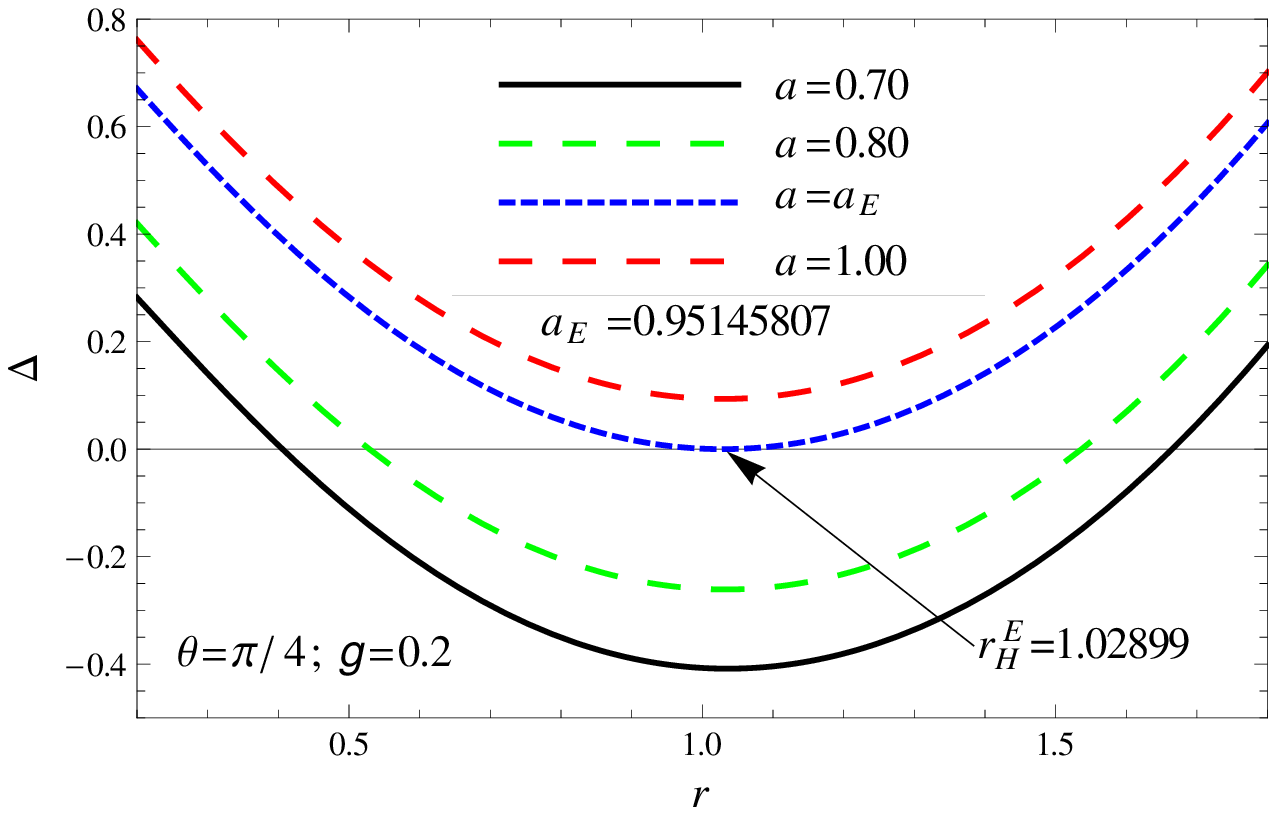}\hspace{-0.7cm}
	   &\includegraphics[scale=0.62]{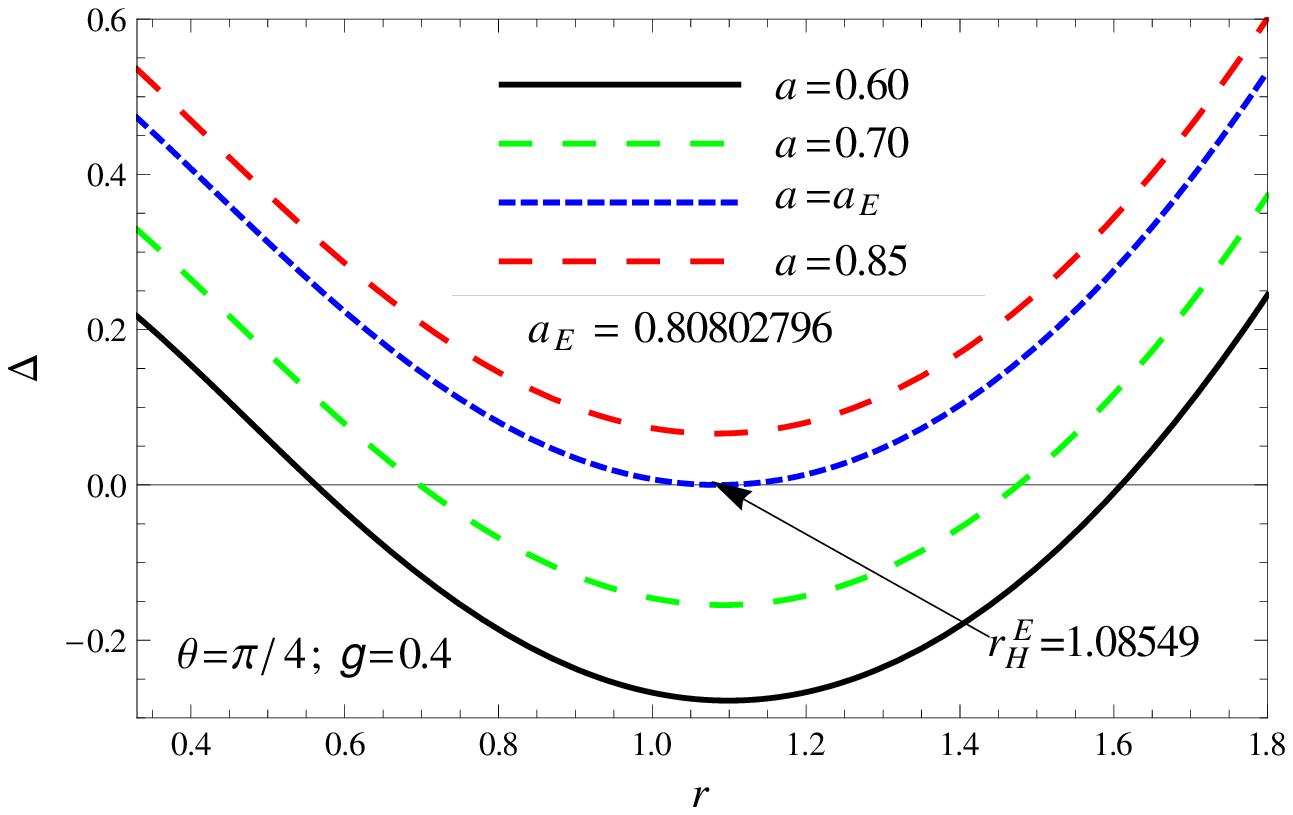}\\
		\end{tabular}
	\caption{Plot showing the behavior of $\Delta$ vs $r$ for fixed values of $\zeta=2$, $\lambda=3$, $\theta=\pi/4$ and $M=1$ by varying $a$. Case $a=a_E$ corresponds to an extremal black hole}\label{fig1}
\end{figure*}

\begin{figure*}
	\begin{tabular}{c c}
		\includegraphics[scale=0.62]{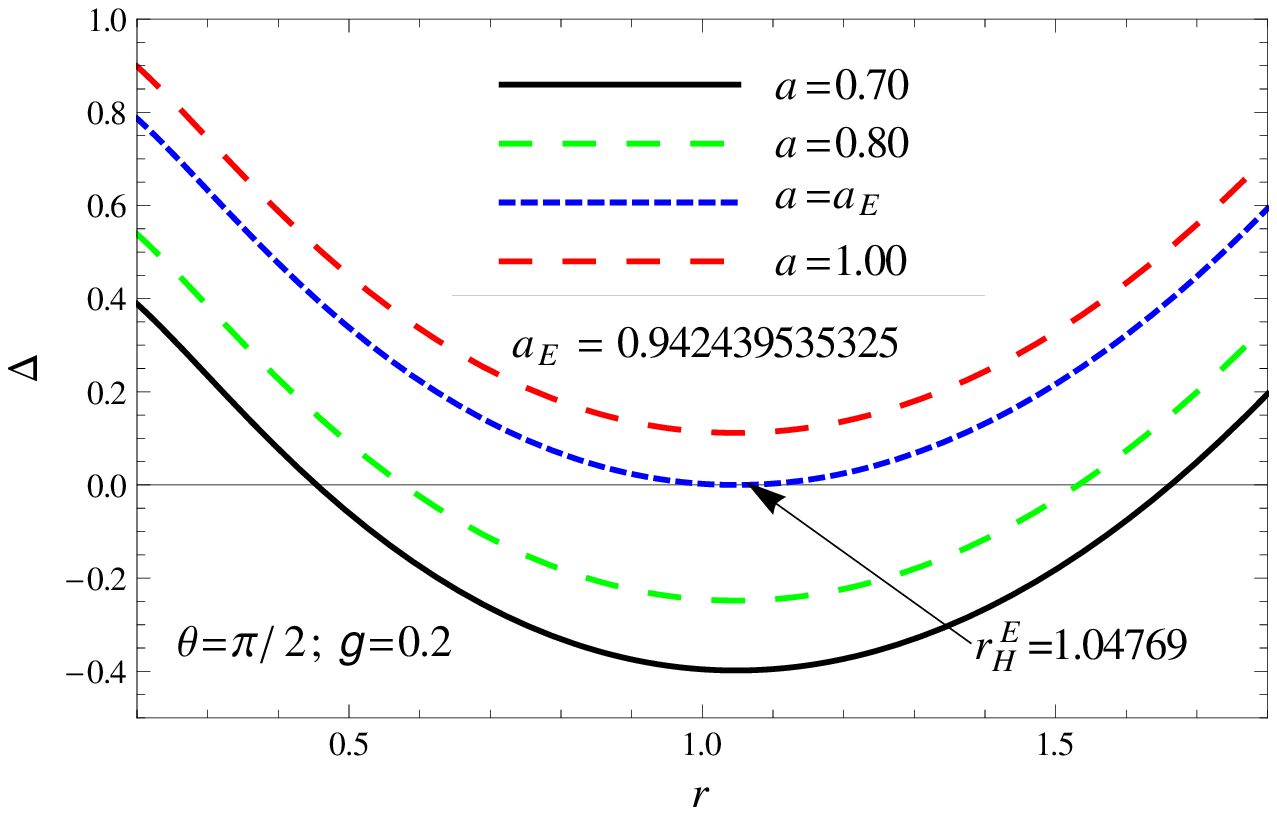}\hspace{-0.7cm}
	   &\includegraphics[scale=0.62]{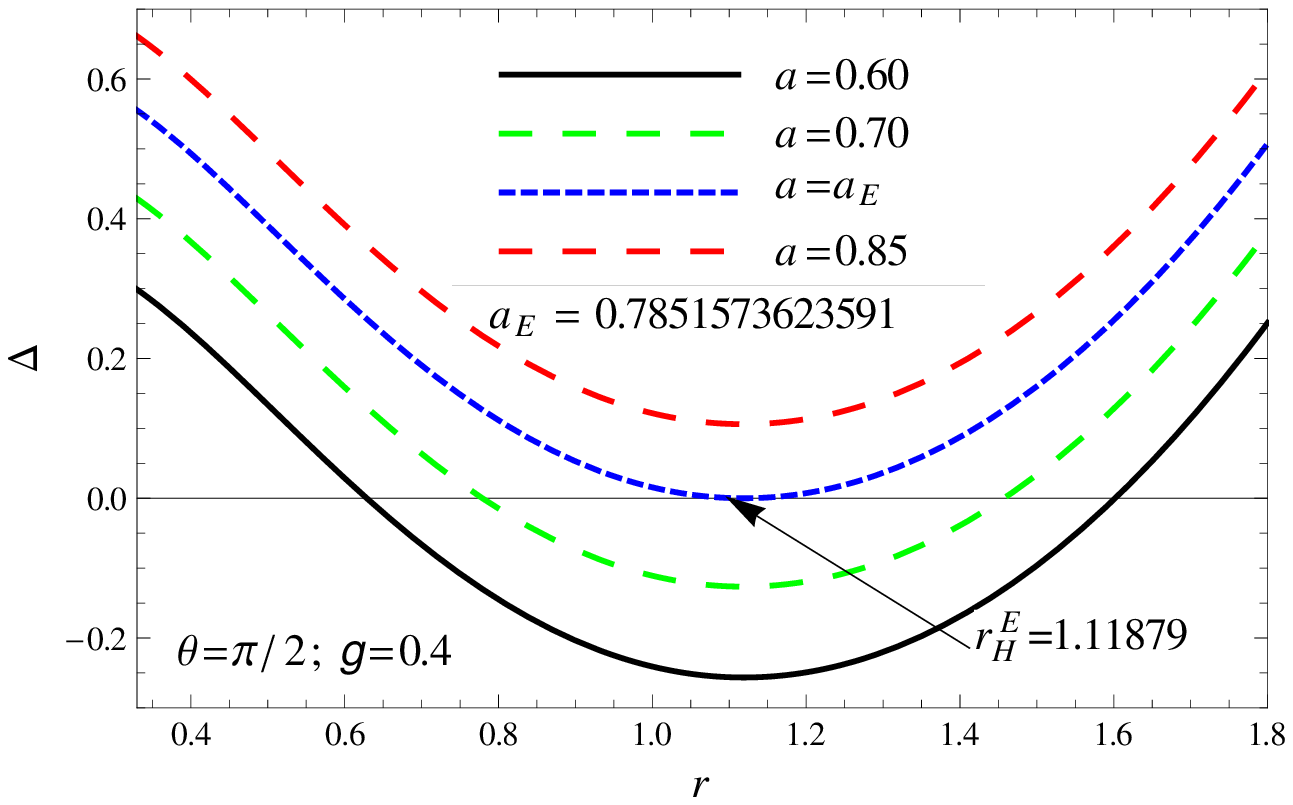}\\
		\includegraphics[scale=0.62]{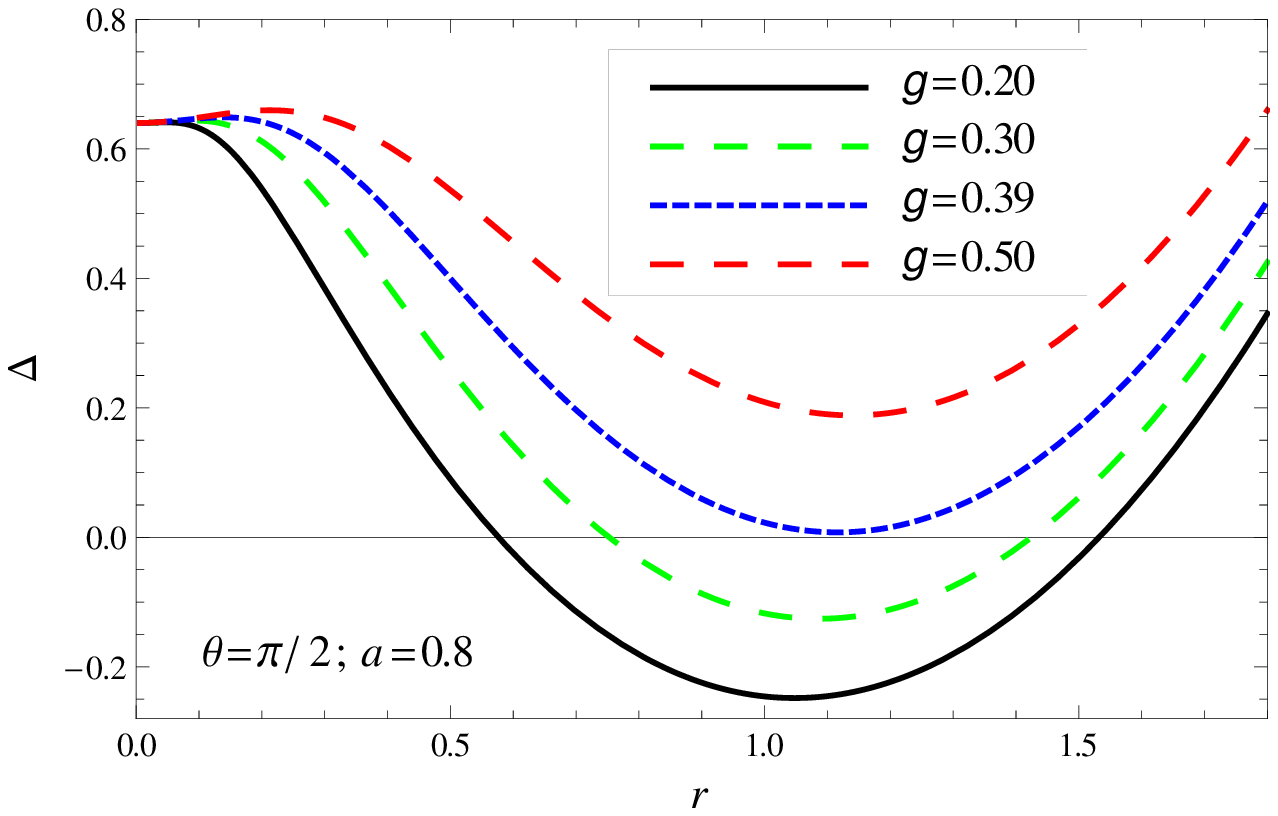}\hspace{-0.7cm}
	   &\includegraphics[scale=0.62]{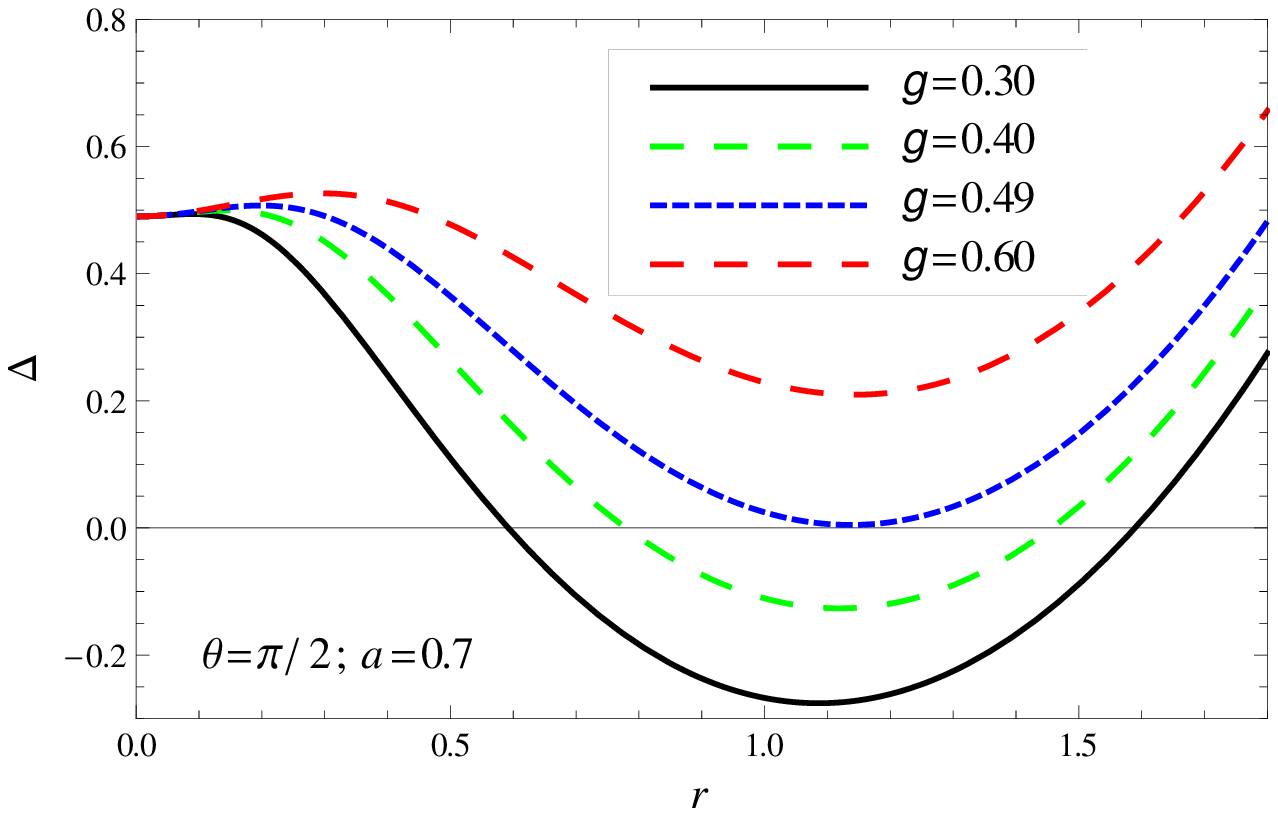}
	\end{tabular}
	\caption{Plot showing the behavior of $\Delta$ vs $r$ for fixed values of $\theta=\pi/2$ and $M=1$.  Case $a=a_E$ corresponds to an extremal black hole}\label{fig2}
\end{figure*}

\begin{figure}
	\includegraphics[scale=0.5]{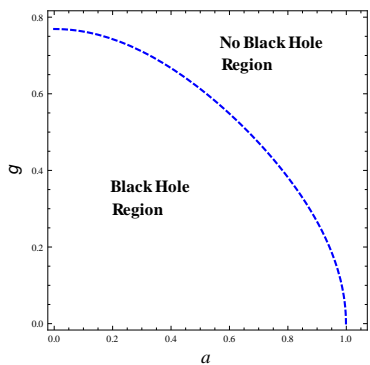}
	\caption{Plot Showing the behavior of spin parameter ($a$) and magnetic charge parameter ($g$) plane of the rotating Bardeen regular metric (contour plot of $\Delta=0$). The blue dashed line is the boundary which separate the black hole region from no black hole region.}\label{fig3}
\end{figure}

\begin{figure*}
	\begin{tabular}{c c c c}
		\includegraphics[scale=0.62]{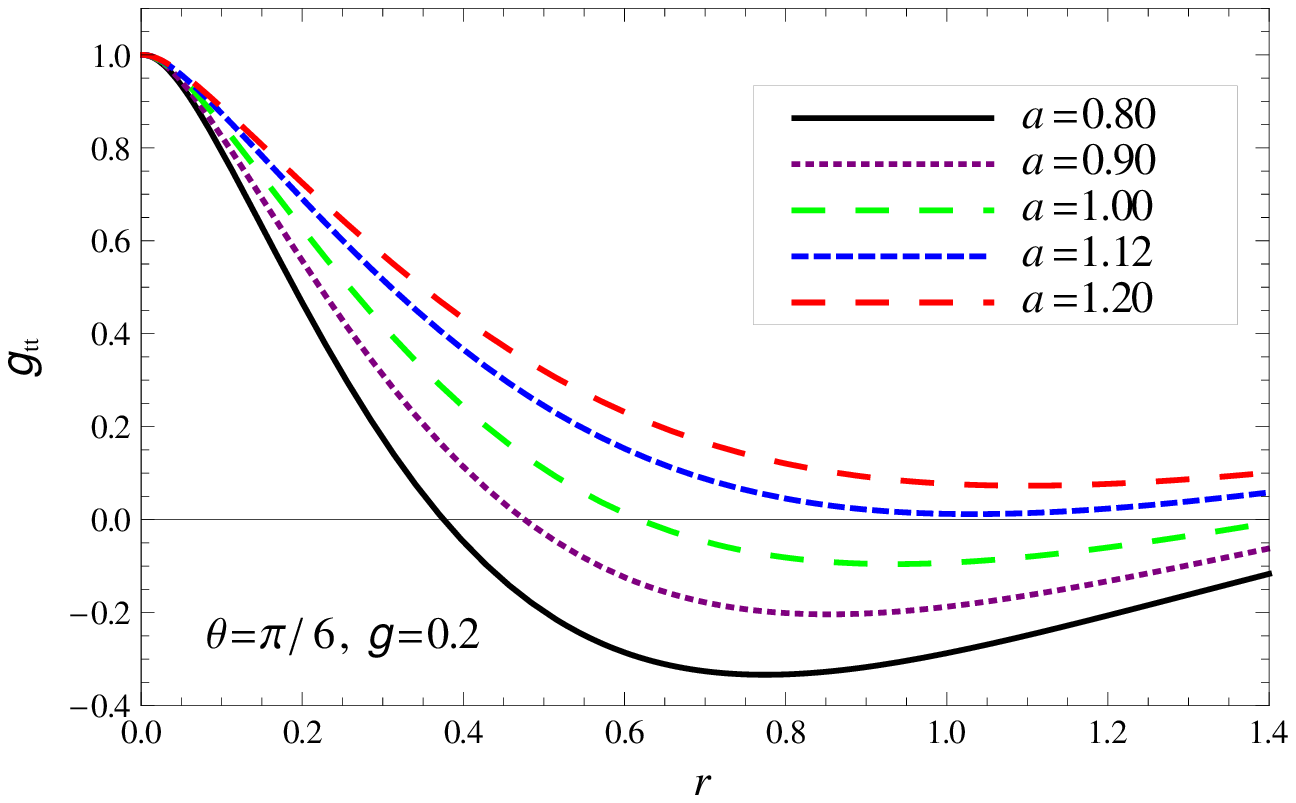}\hspace{-0.7cm}
	   &\includegraphics[scale=0.62]{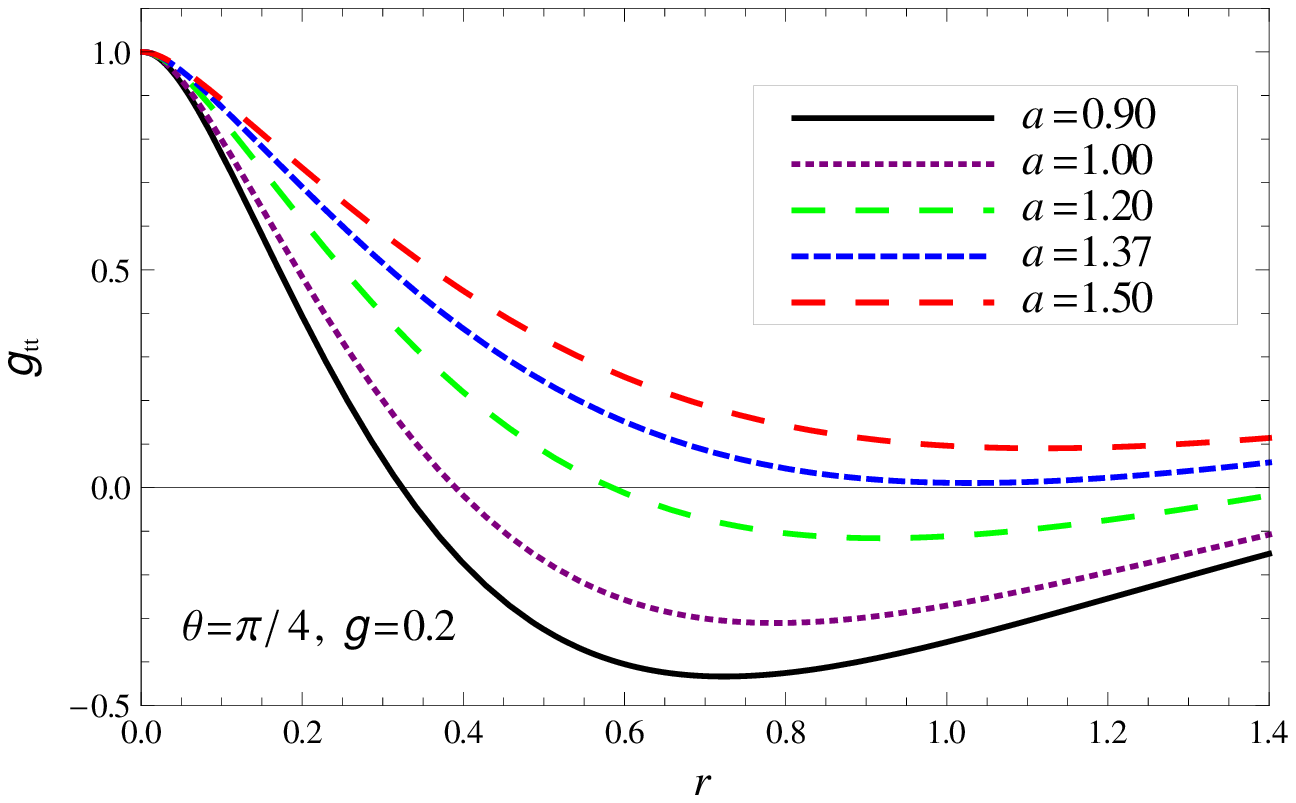}\\
		\includegraphics[scale=0.62]{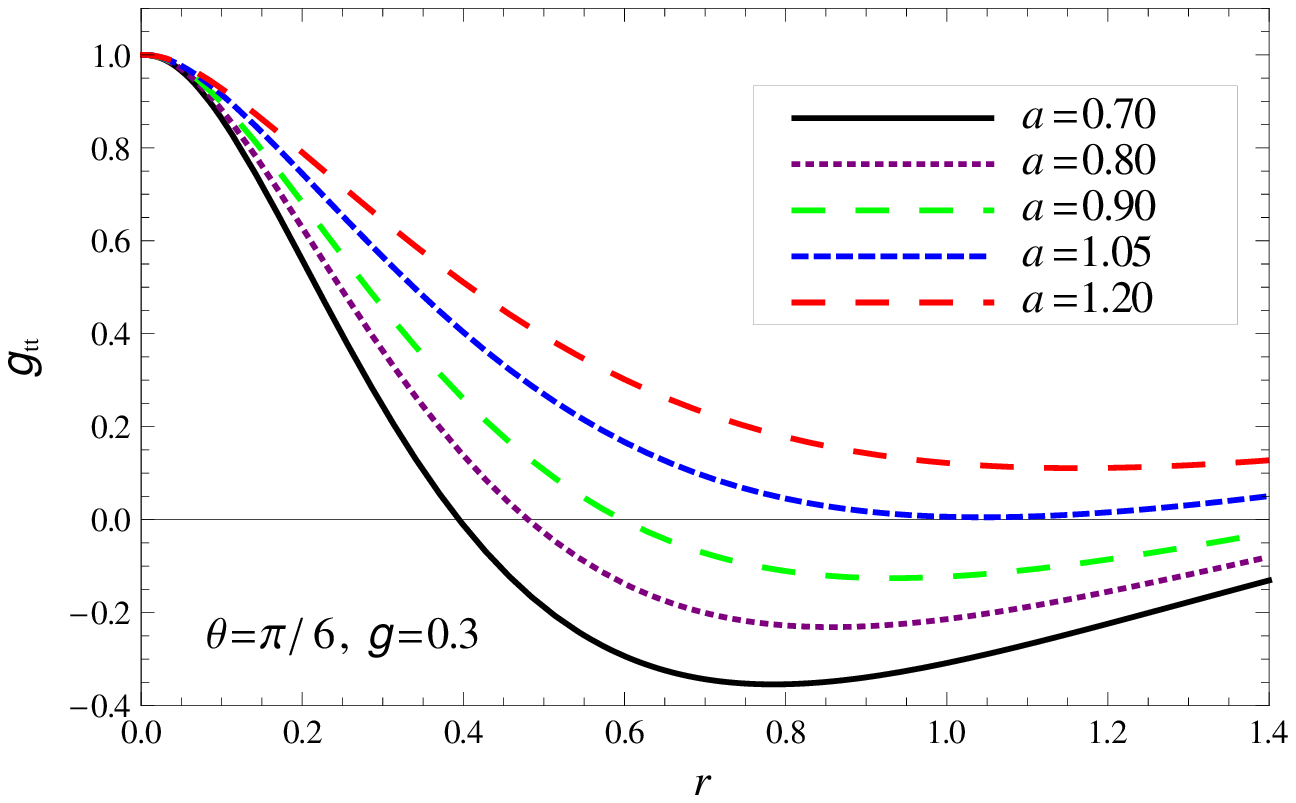}\hspace{-0.7cm}
	   &\includegraphics[scale=0.62]{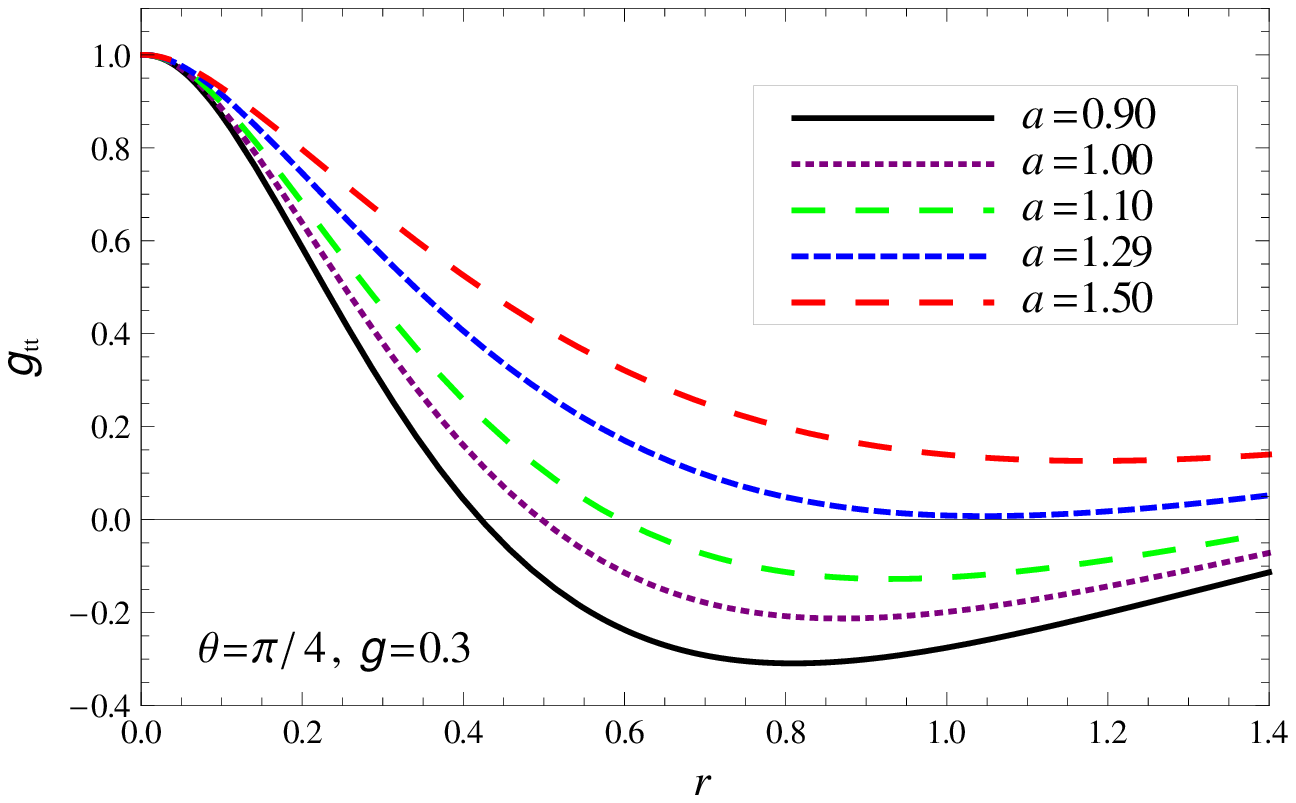}\\
		\includegraphics[scale=0.62]{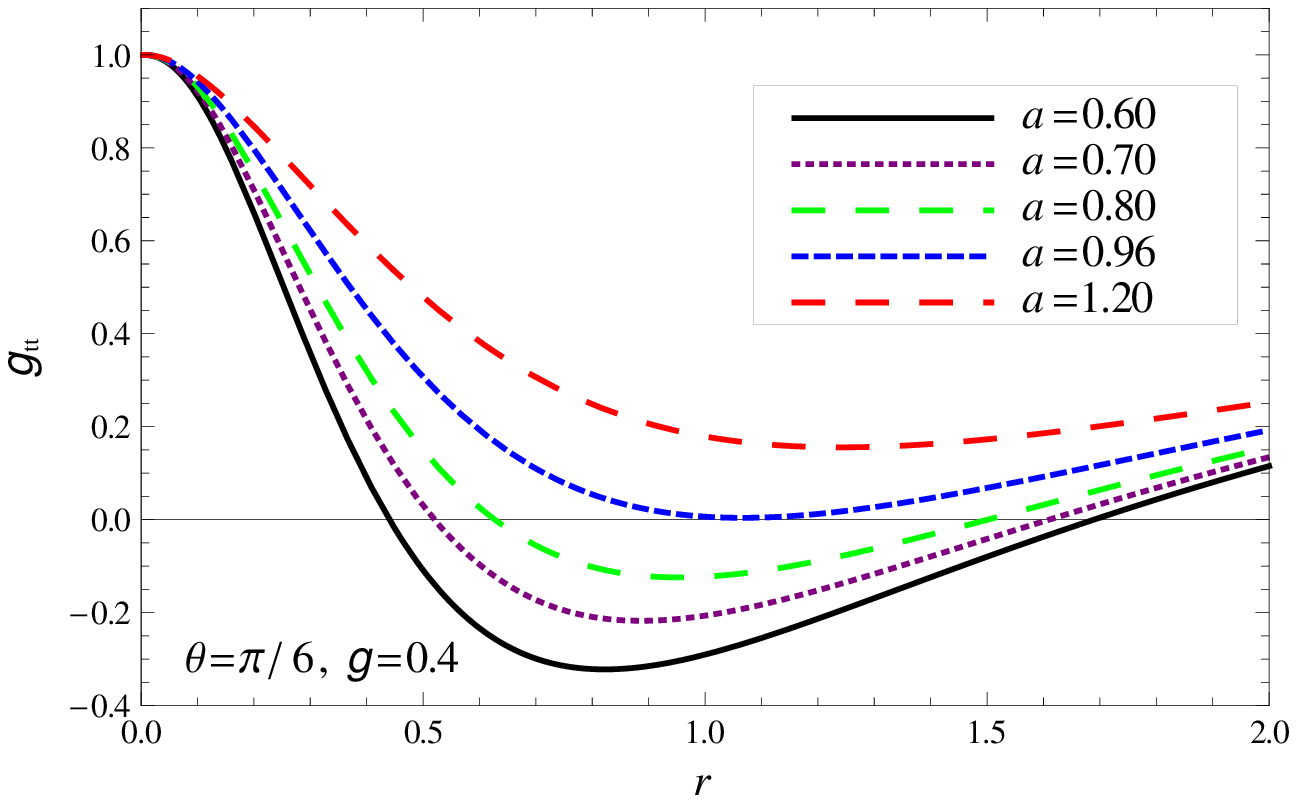}\hspace{-0.7cm}
	   &\includegraphics[scale=0.62]{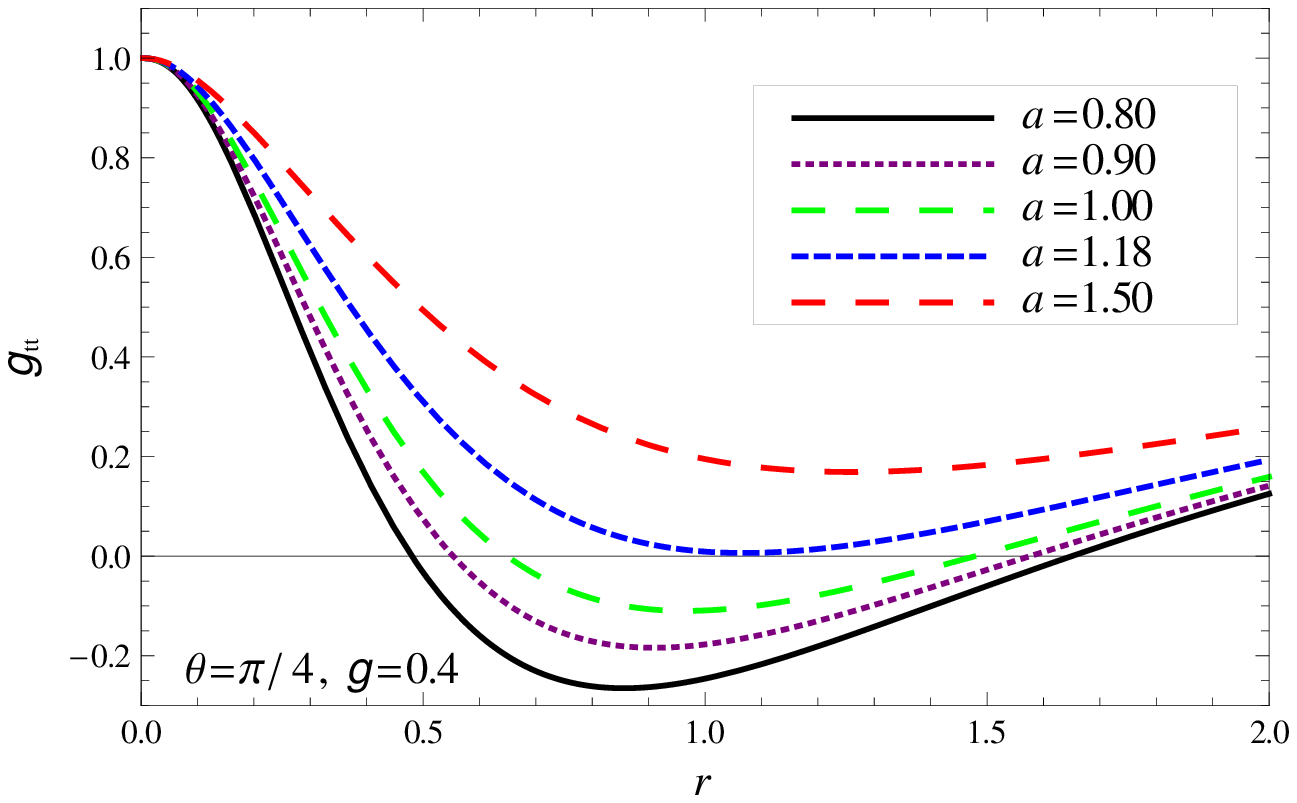}
	\end{tabular}
	\caption{Plot showing the variation of infinite redshift surface with the parameter $a$, $g$ and $\theta$.}\label{fig4}
\end{figure*}

\begin{figure*}
	\begin{tabular}{c c c c}
		\includegraphics[scale=0.42]{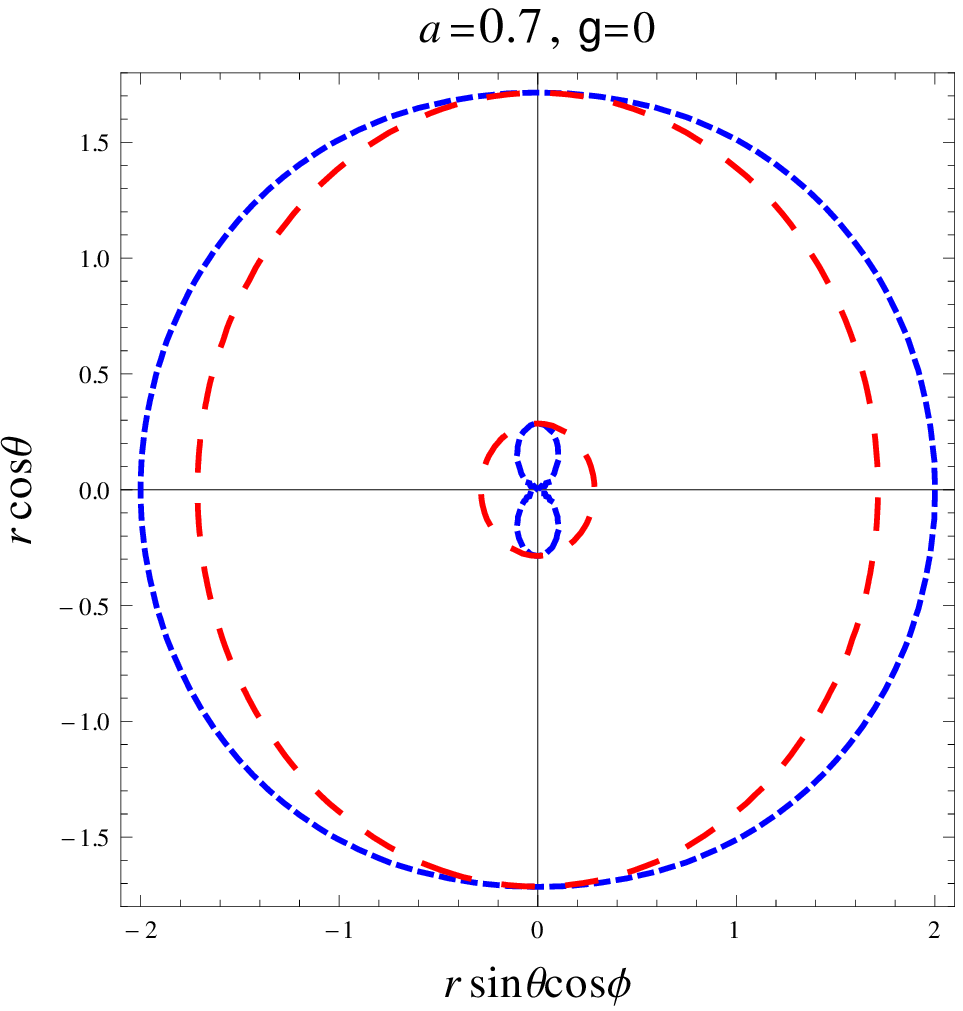}\hspace{-0.2cm}
		\includegraphics[scale=0.42]{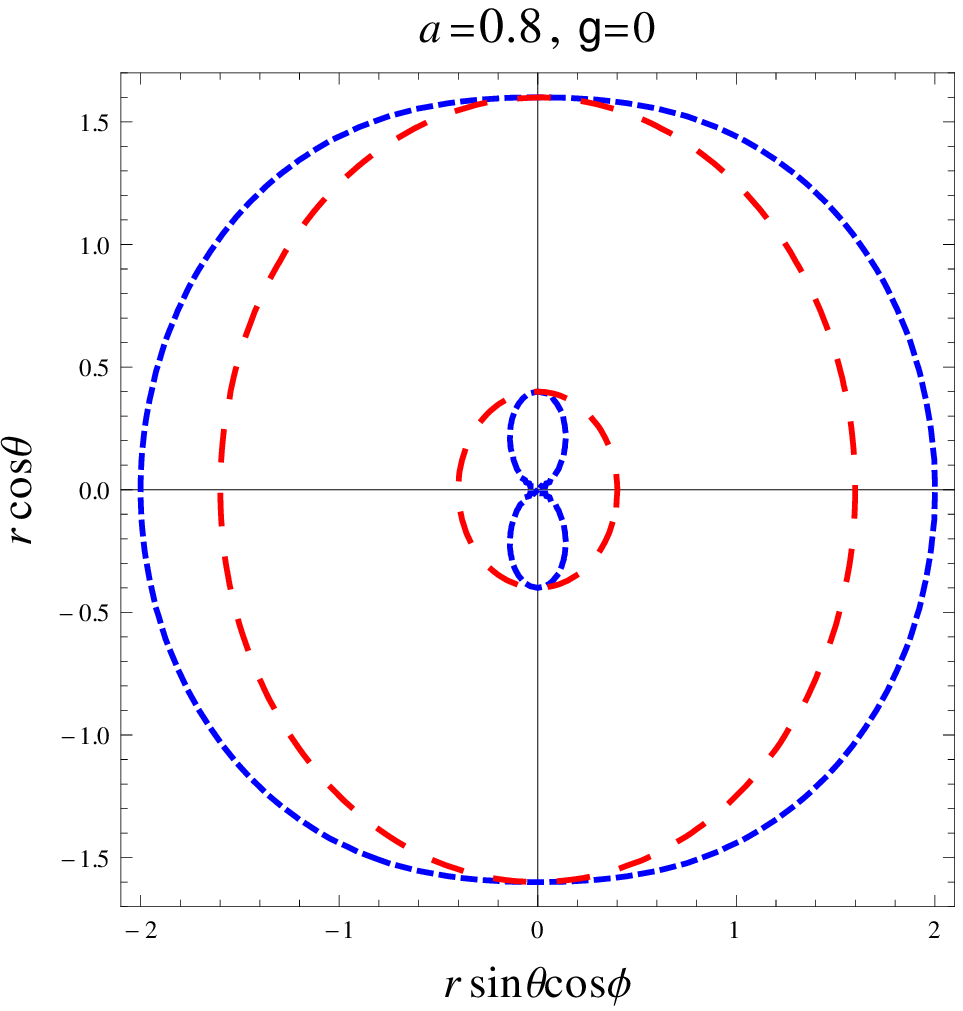}\hspace{-0.2cm}
		\includegraphics[scale=0.42]{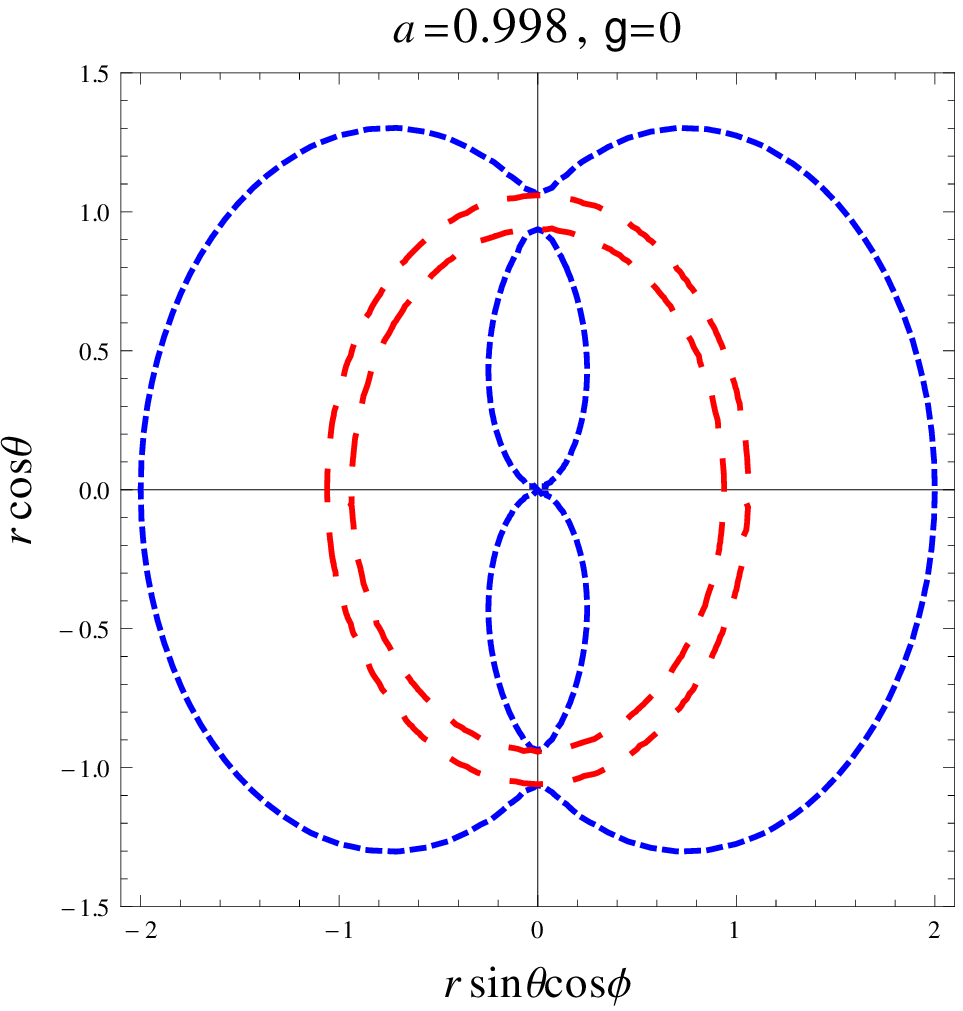}\hspace{-0.2cm}
	   &\includegraphics[scale=0.42]{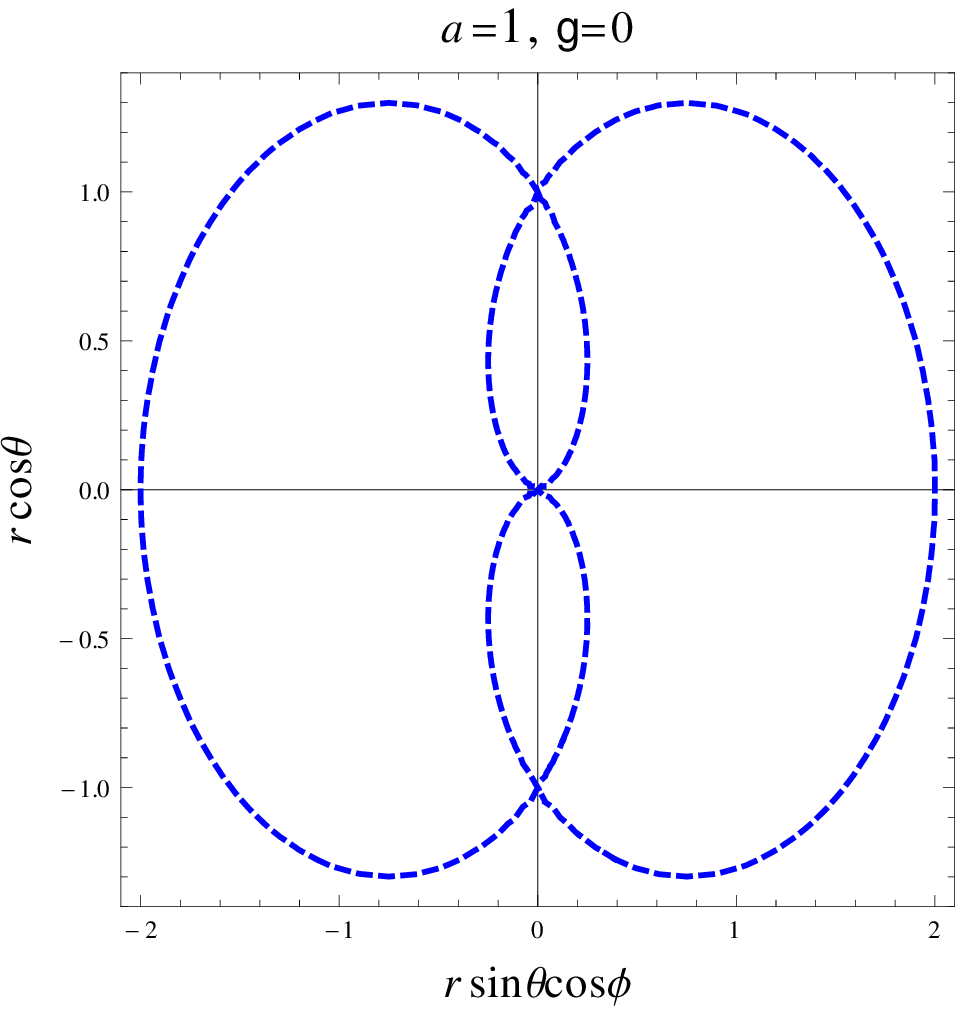}\\
		\includegraphics[scale=0.42]{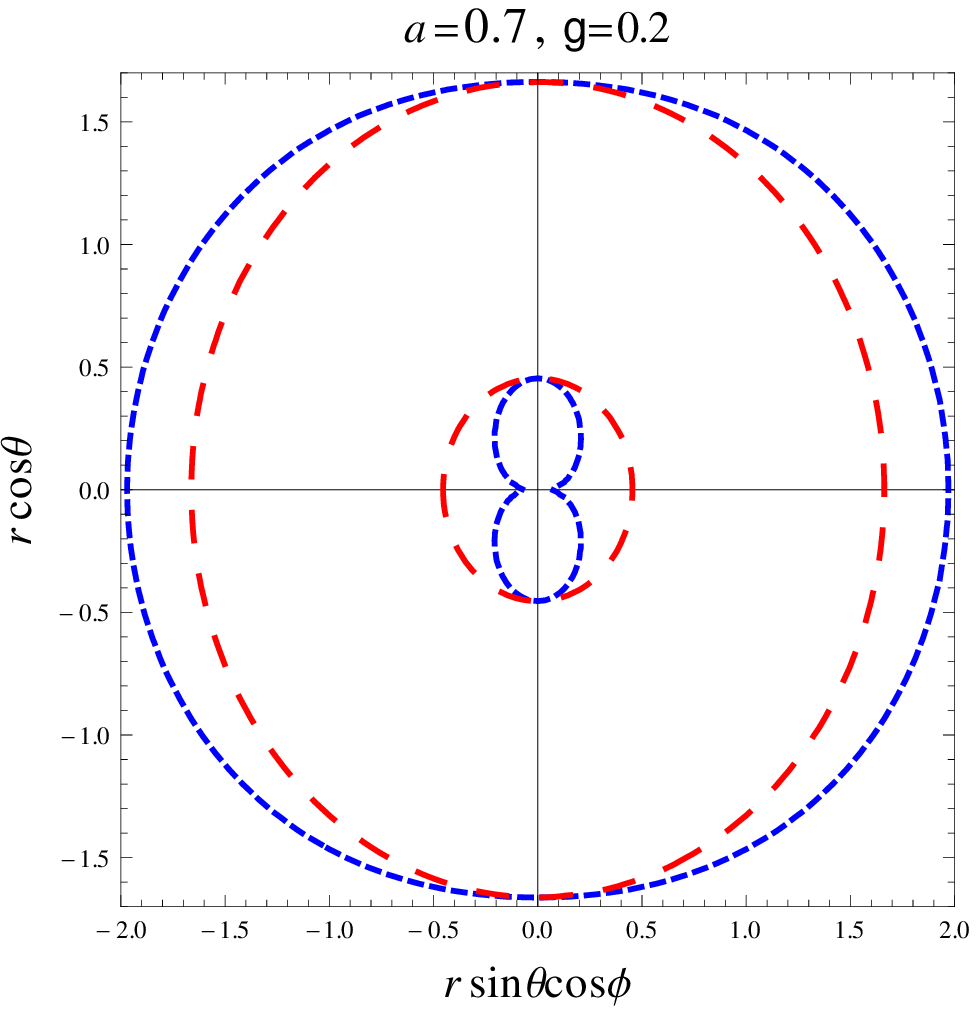}\hspace{-0.2cm}
		\includegraphics[scale=0.42]{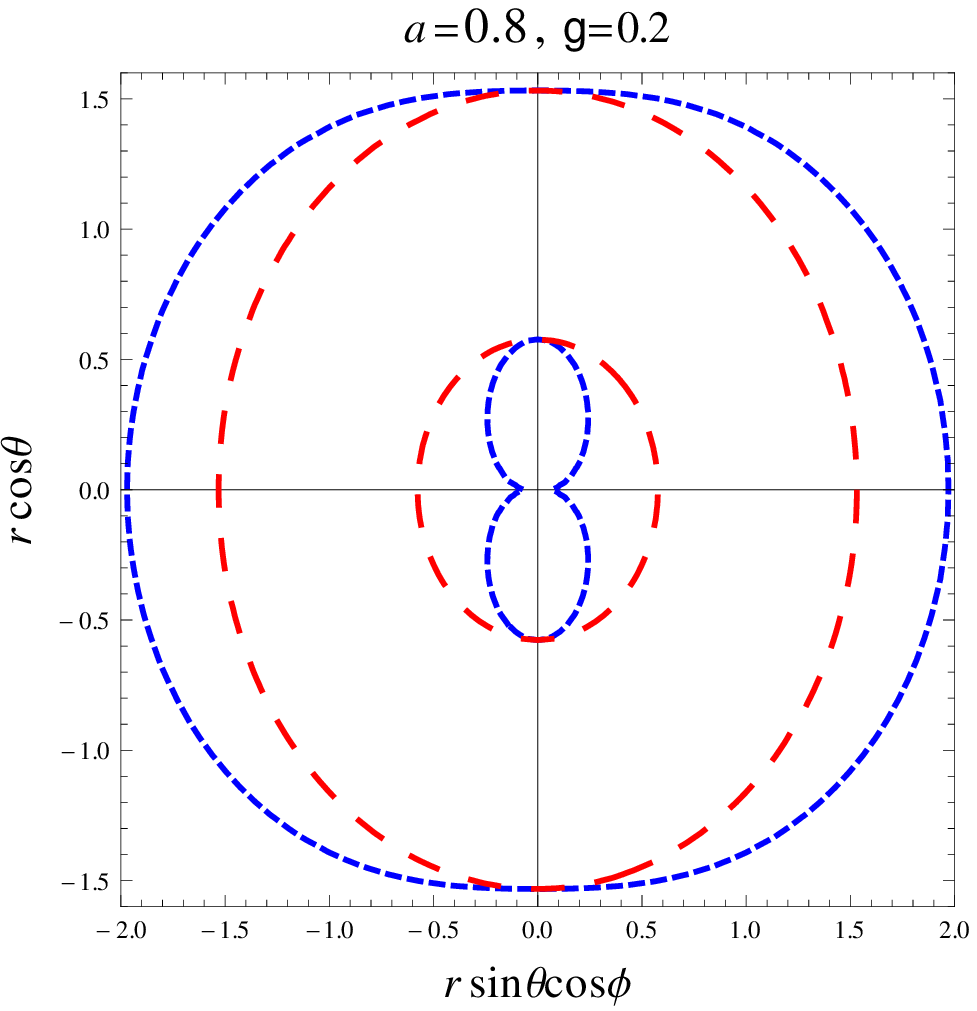}\hspace{-0.2cm}
		\includegraphics[scale=0.42]{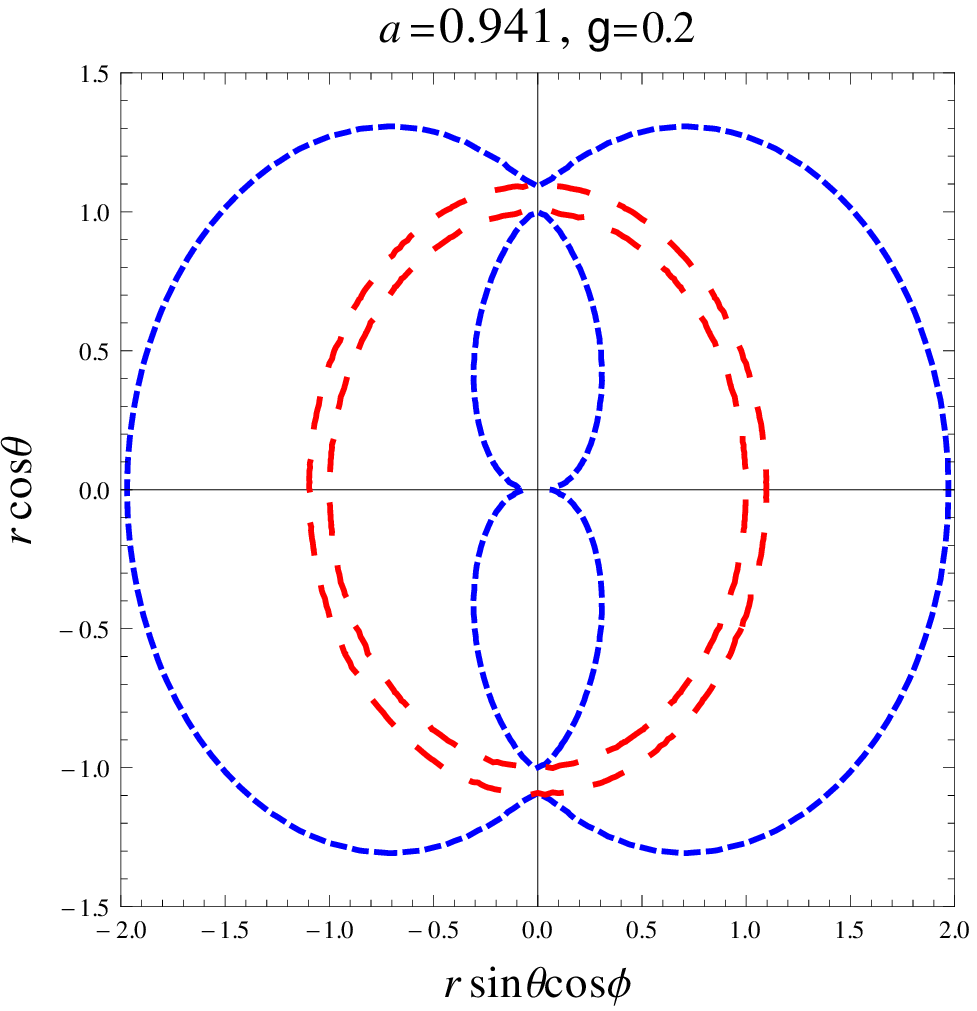}\hspace{-0.2cm}
	   &\includegraphics[scale=0.42]{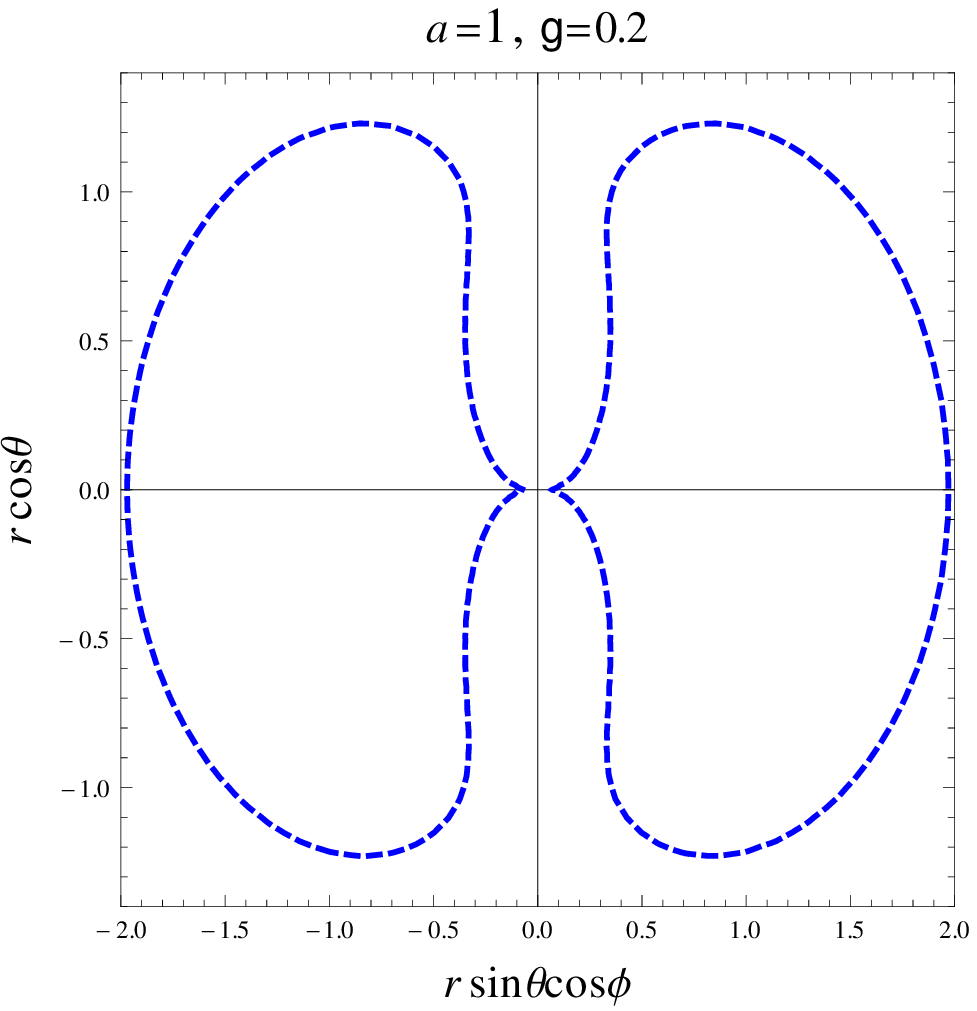}\\
		\includegraphics[scale=0.42]{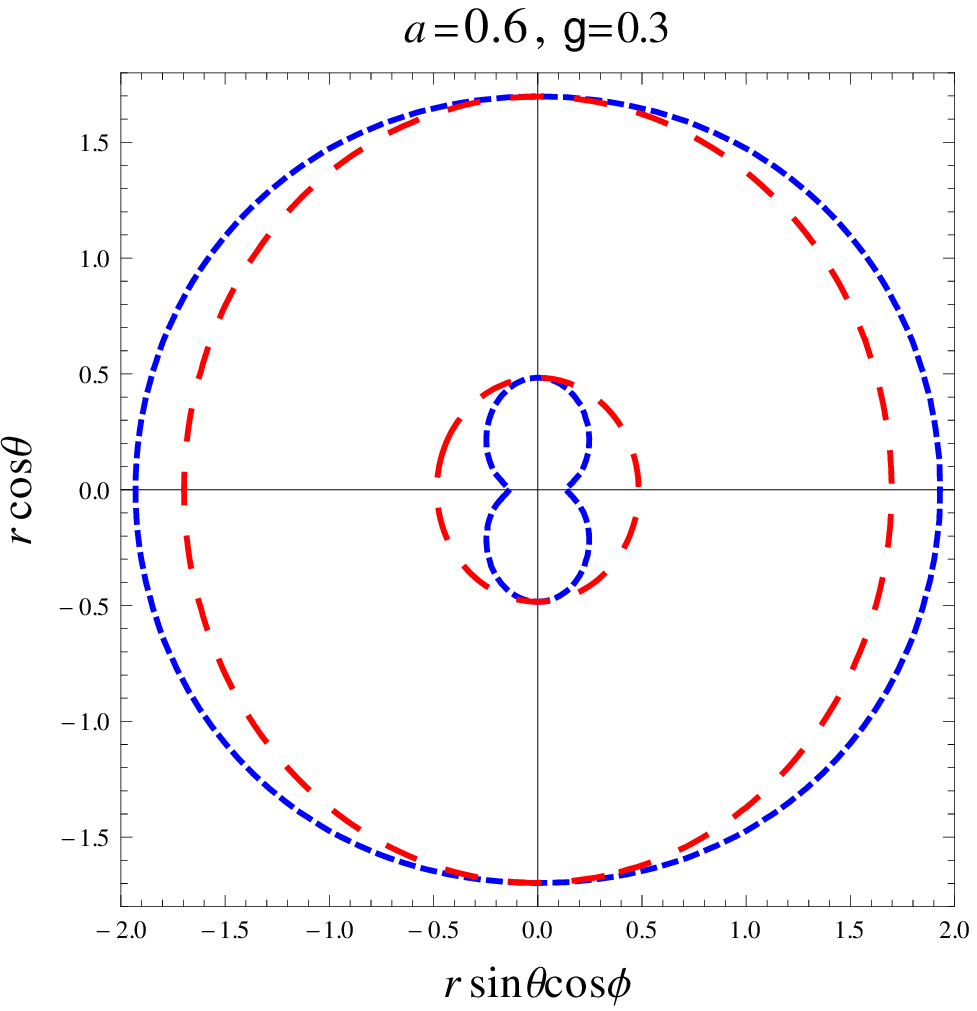}\hspace{-0.2cm}
		\includegraphics[scale=0.42]{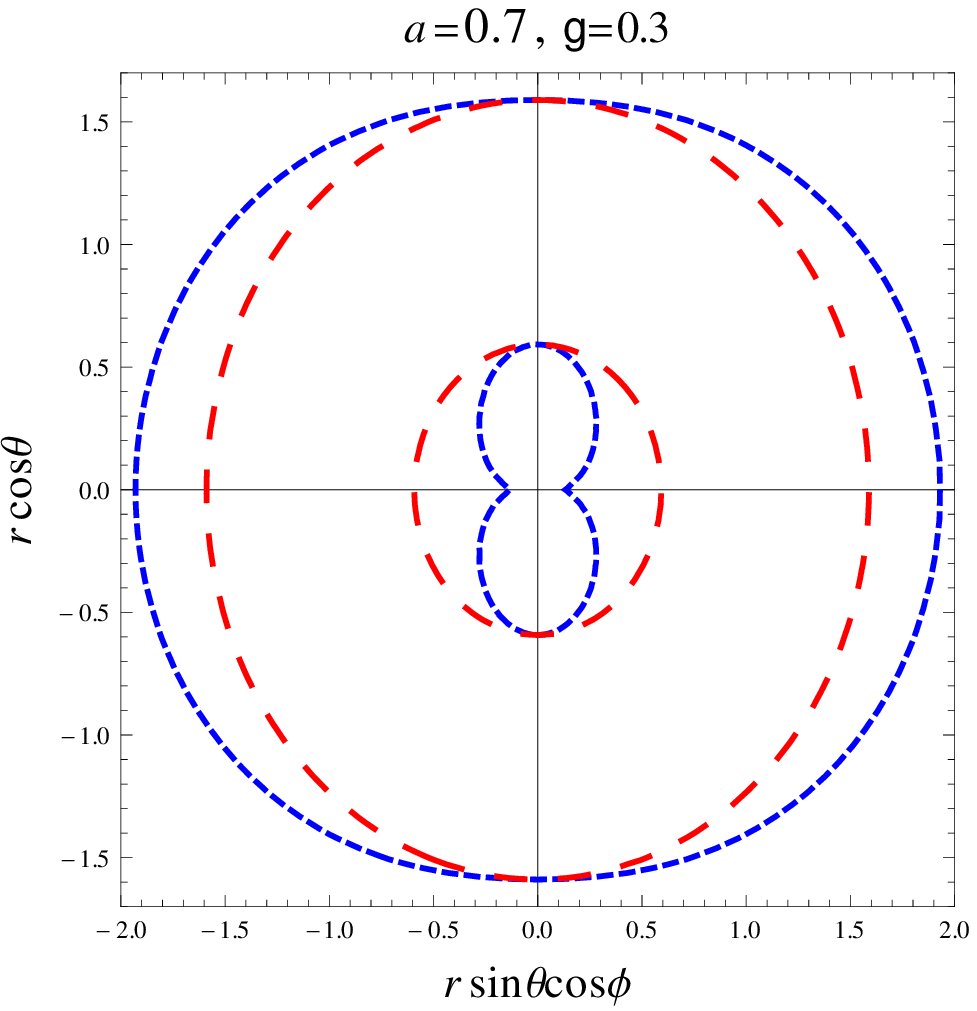}\hspace{-0.2cm}
		\includegraphics[scale=0.42]{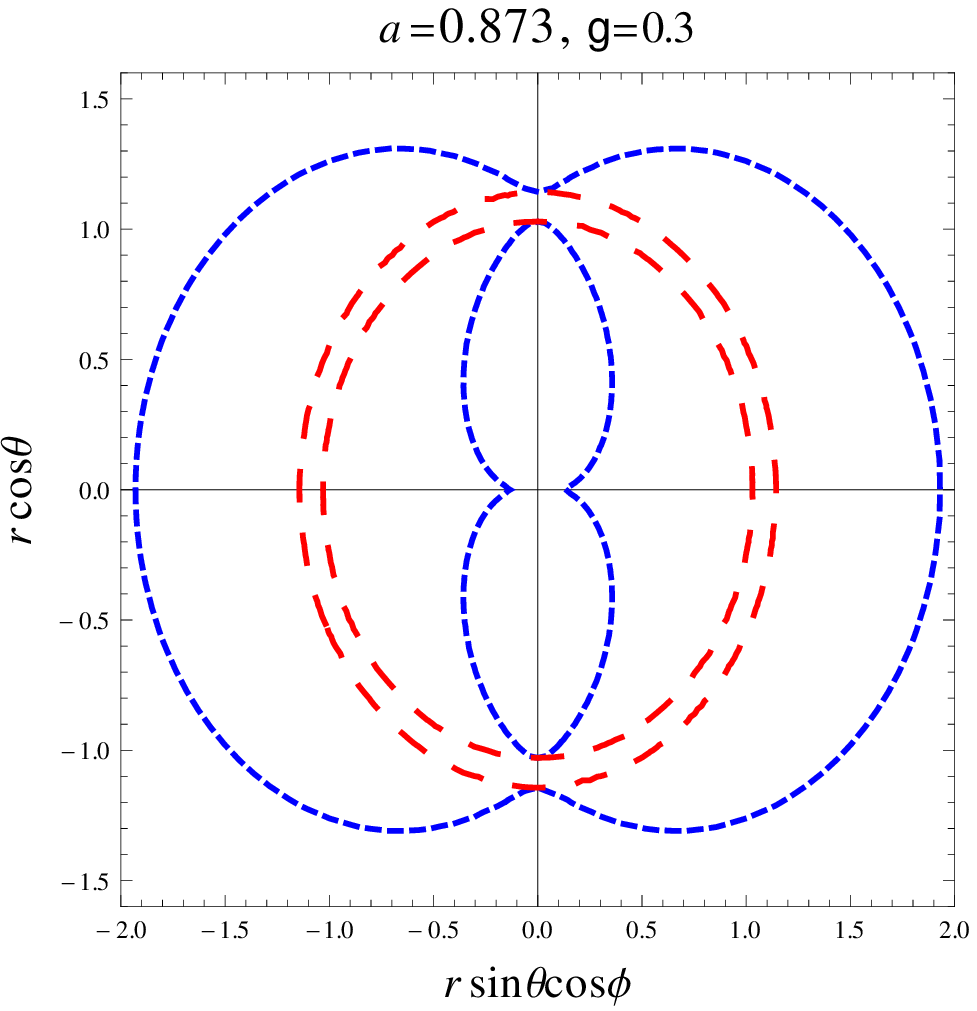}\hspace{-0.2cm}
	   &\includegraphics[scale=0.42]{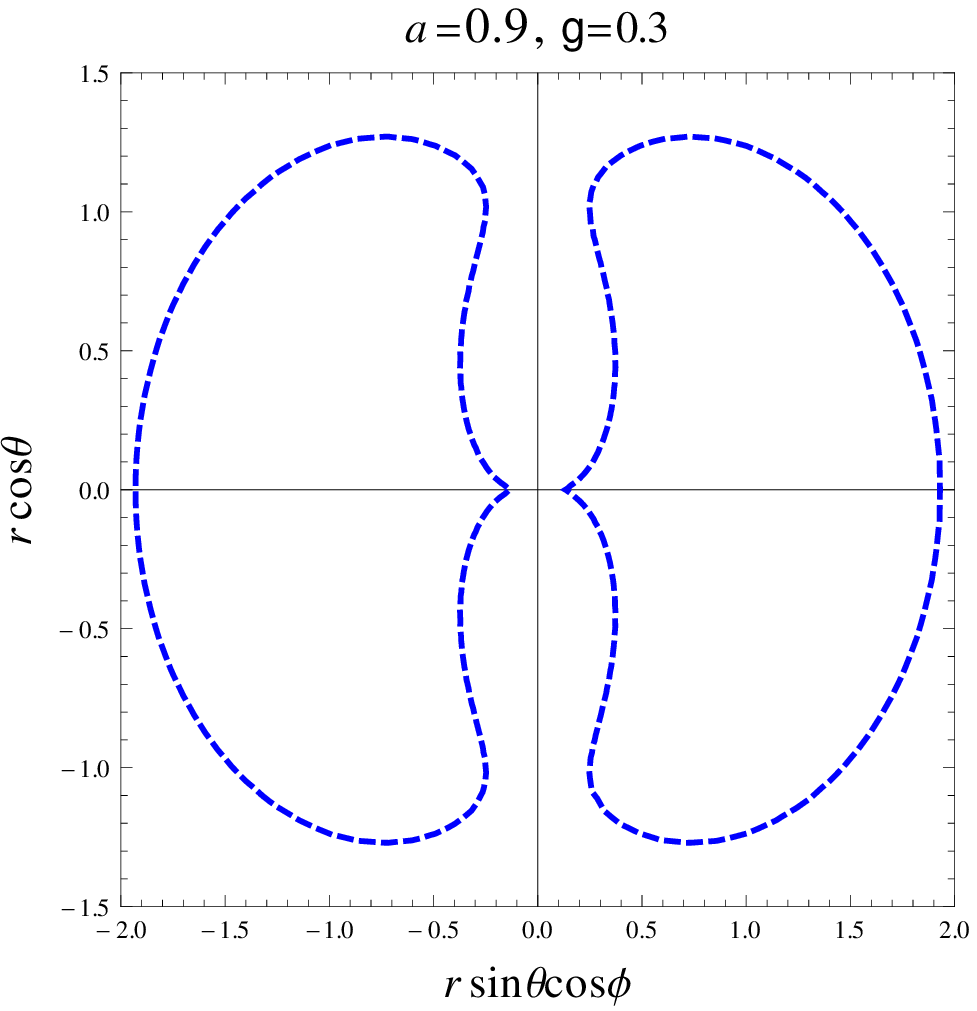}\\
		\includegraphics[scale=0.42]{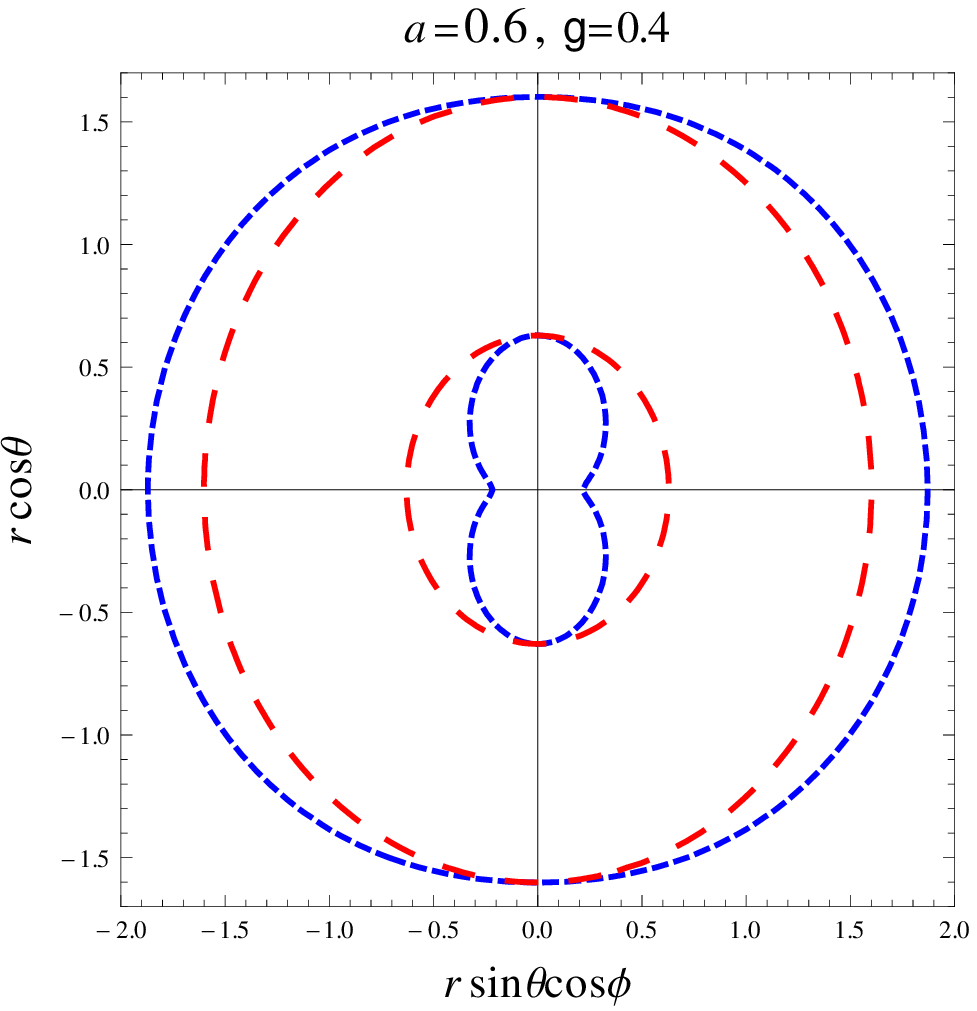}\hspace{-0.2cm}
		\includegraphics[scale=0.42]{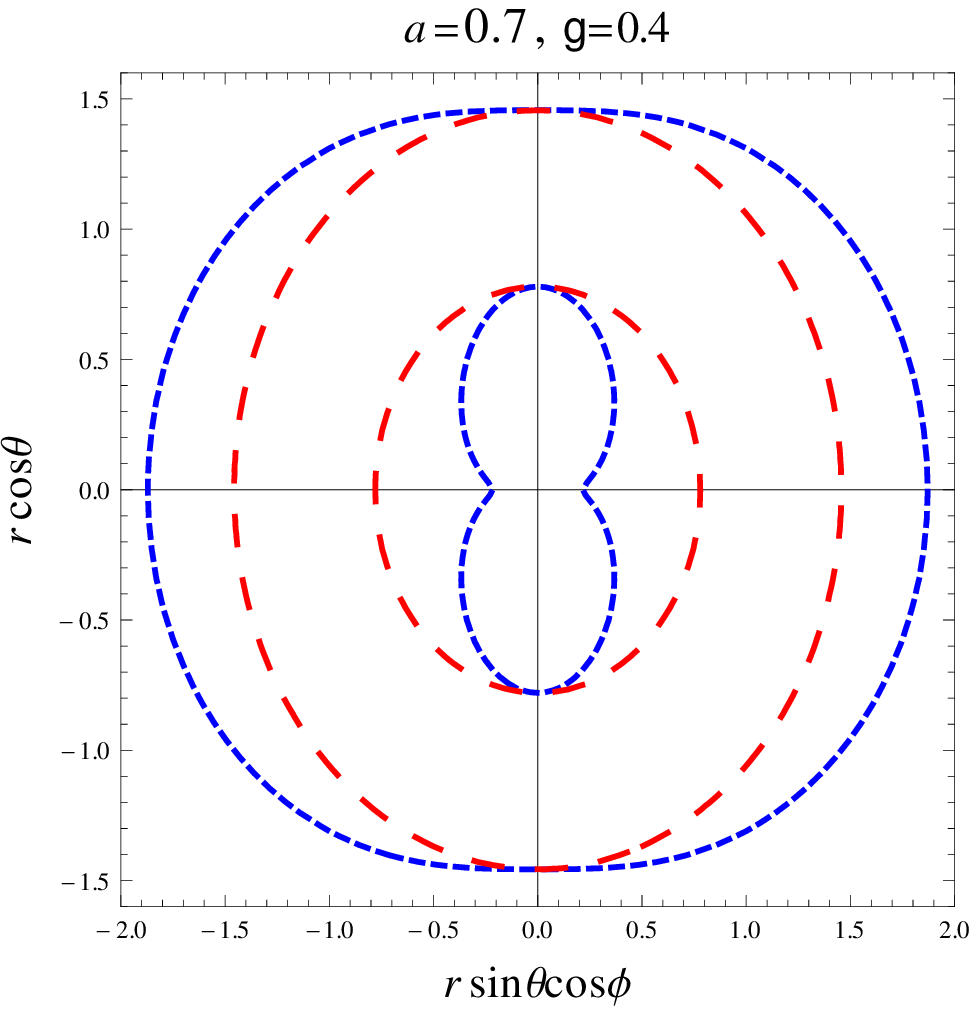}\hspace{-0.2cm}
		\includegraphics[scale=0.42]{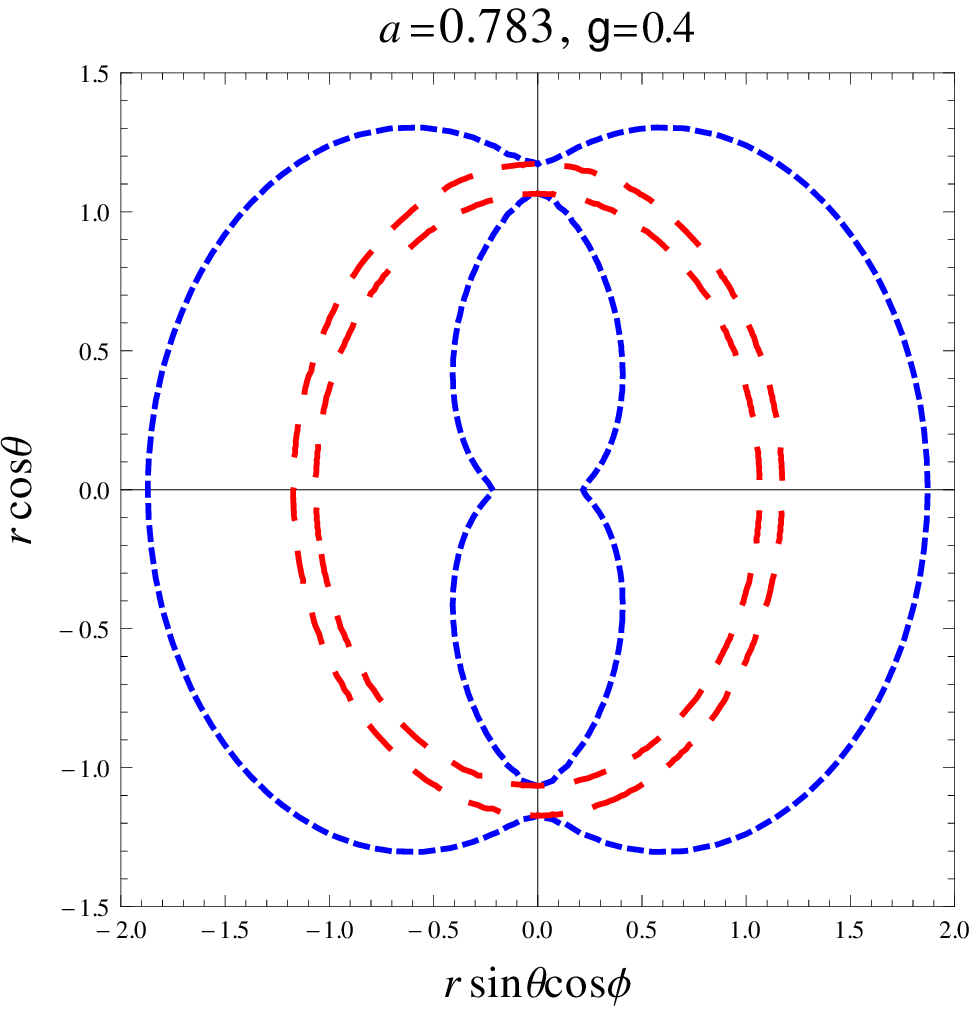}\hspace{-0.2cm}
       &\includegraphics[scale=0.42]{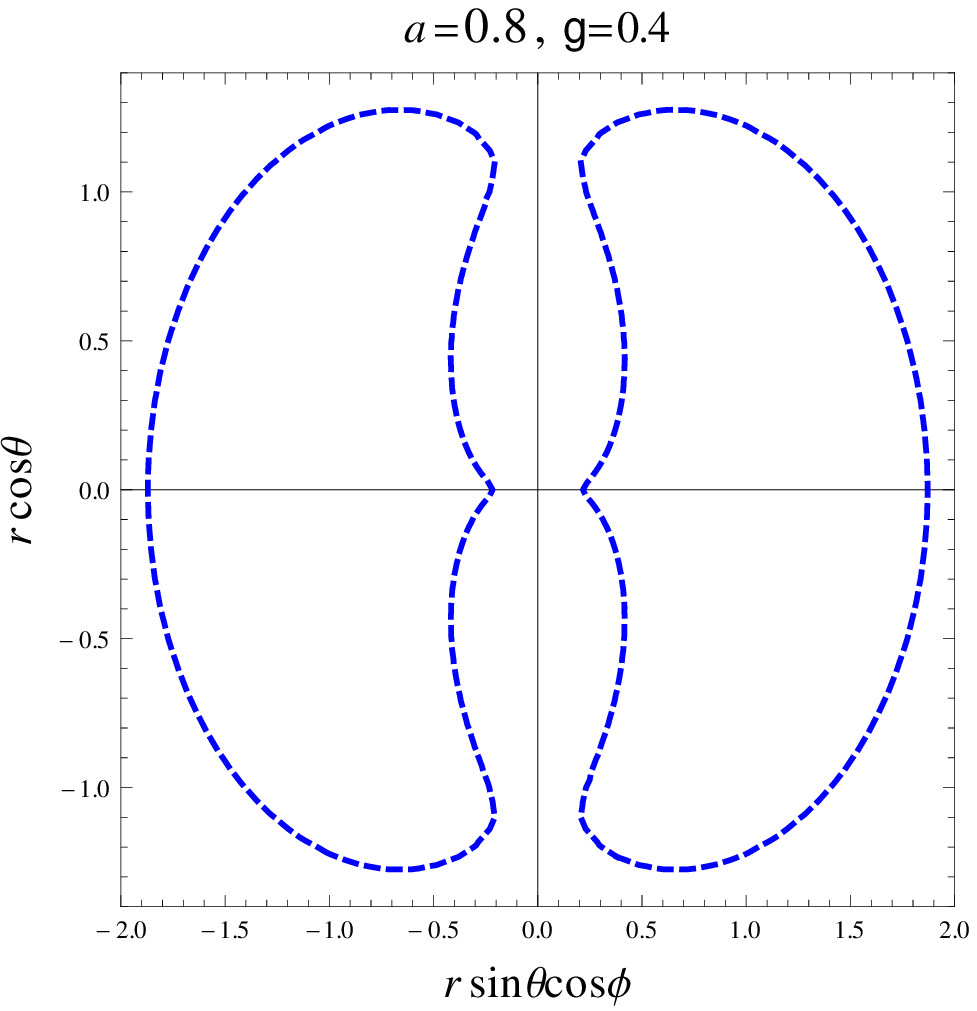}\\
        \includegraphics[scale=0.42]{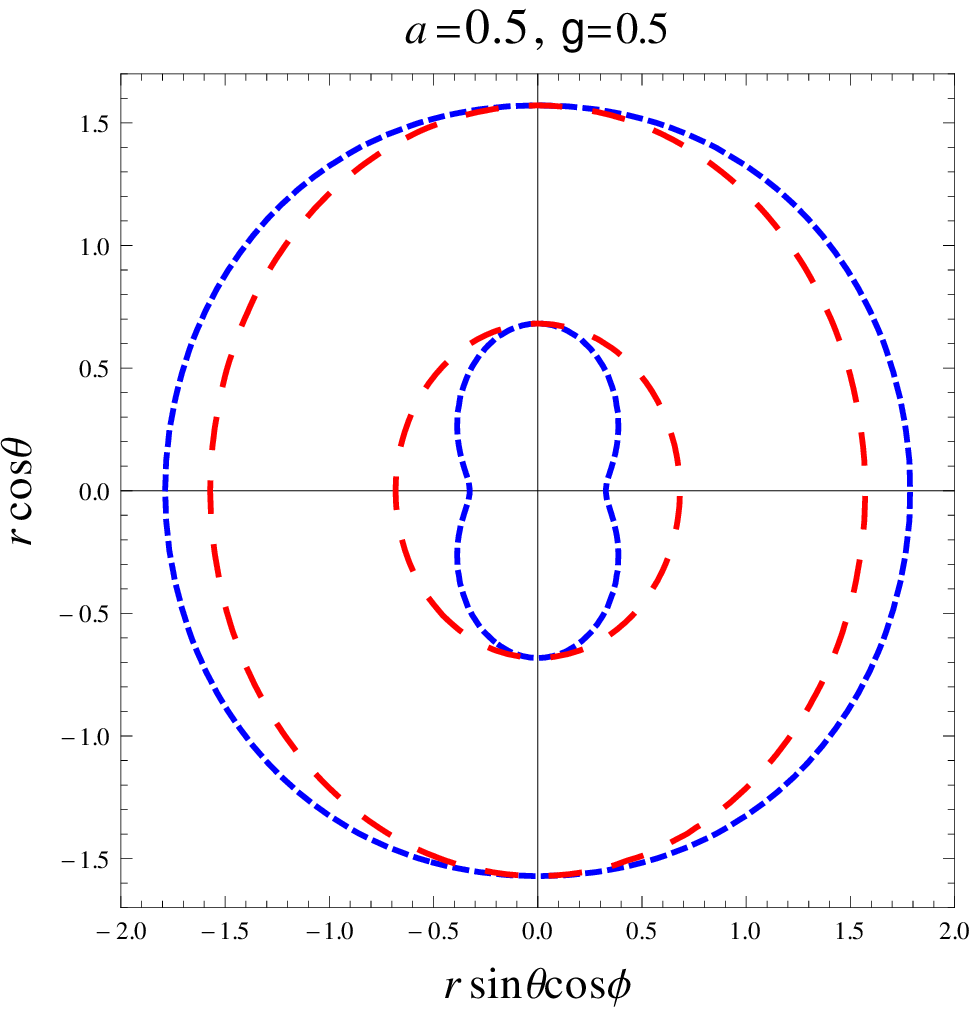}\hspace{-0.2cm}
		\includegraphics[scale=0.42]{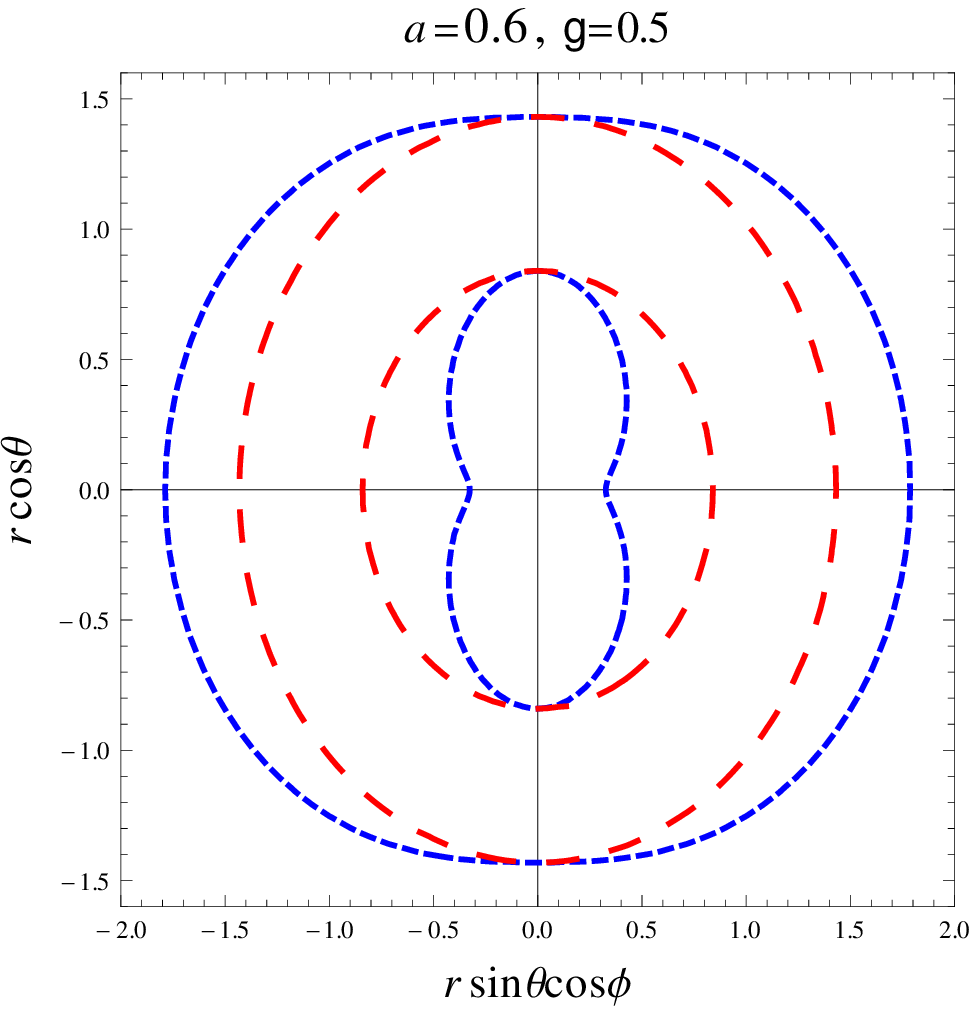}\hspace{-0.2cm}
		\includegraphics[scale=0.42]{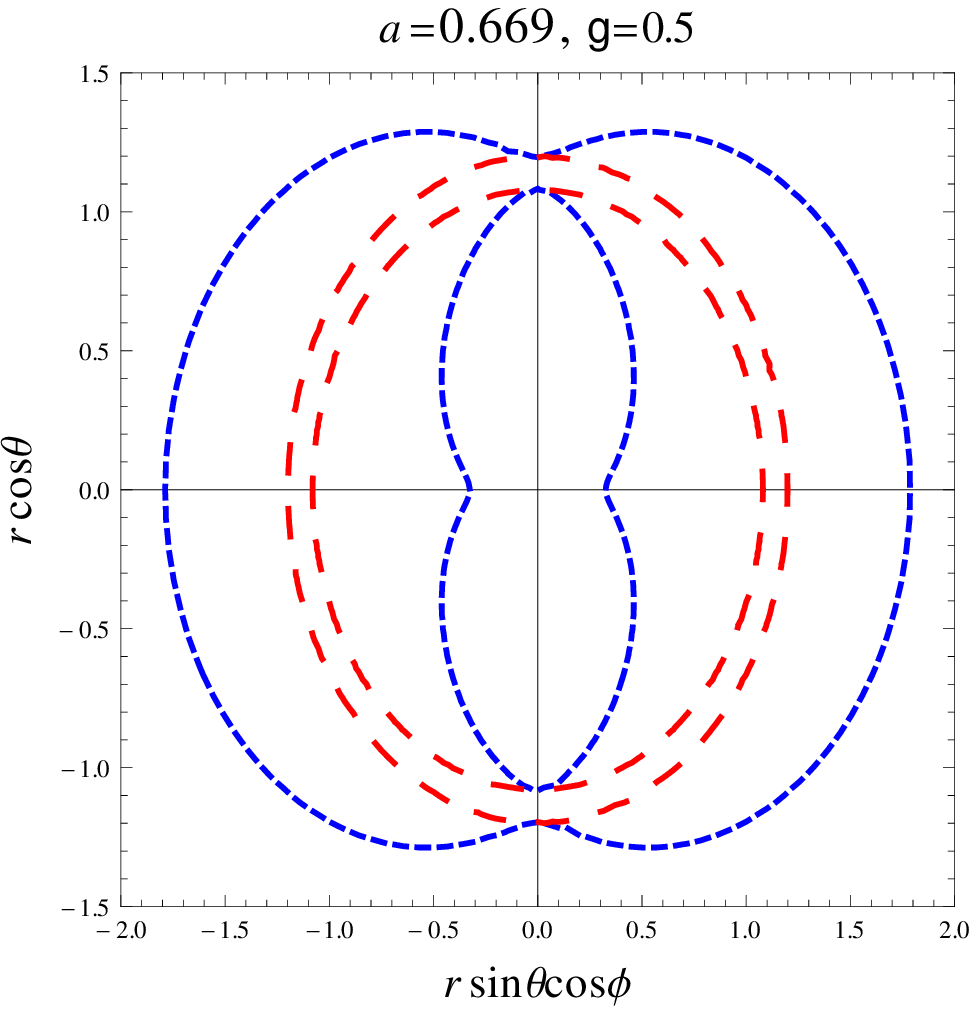}\hspace{-0.2cm}
       &\includegraphics[scale=0.42]{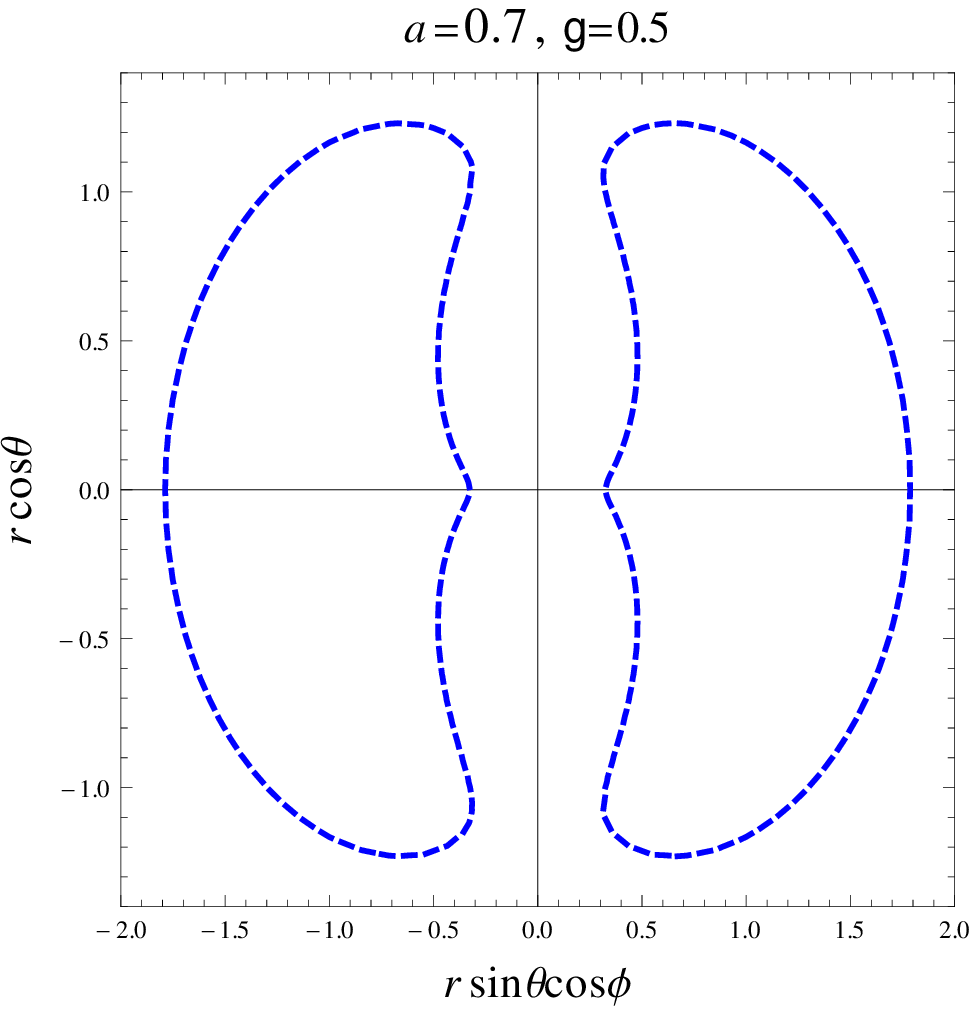}
	\end{tabular}
	\caption{Plot showing the variation of the shape of ergosphere in $xz$-plane with parameter $g$, for different values of $a$, of the rotating Bardeen regular black hole. The blue and the red lines correspond, respectively, to static limit surfaces and horizons. The outer blue line corresponds to the static limit surface, whereas the two red lines correspond to the two horizons.}\label{fig5}
\end{figure*}

\begin{figure*}
	\begin{tabular}{c c c c}
		\includegraphics[scale=0.42]{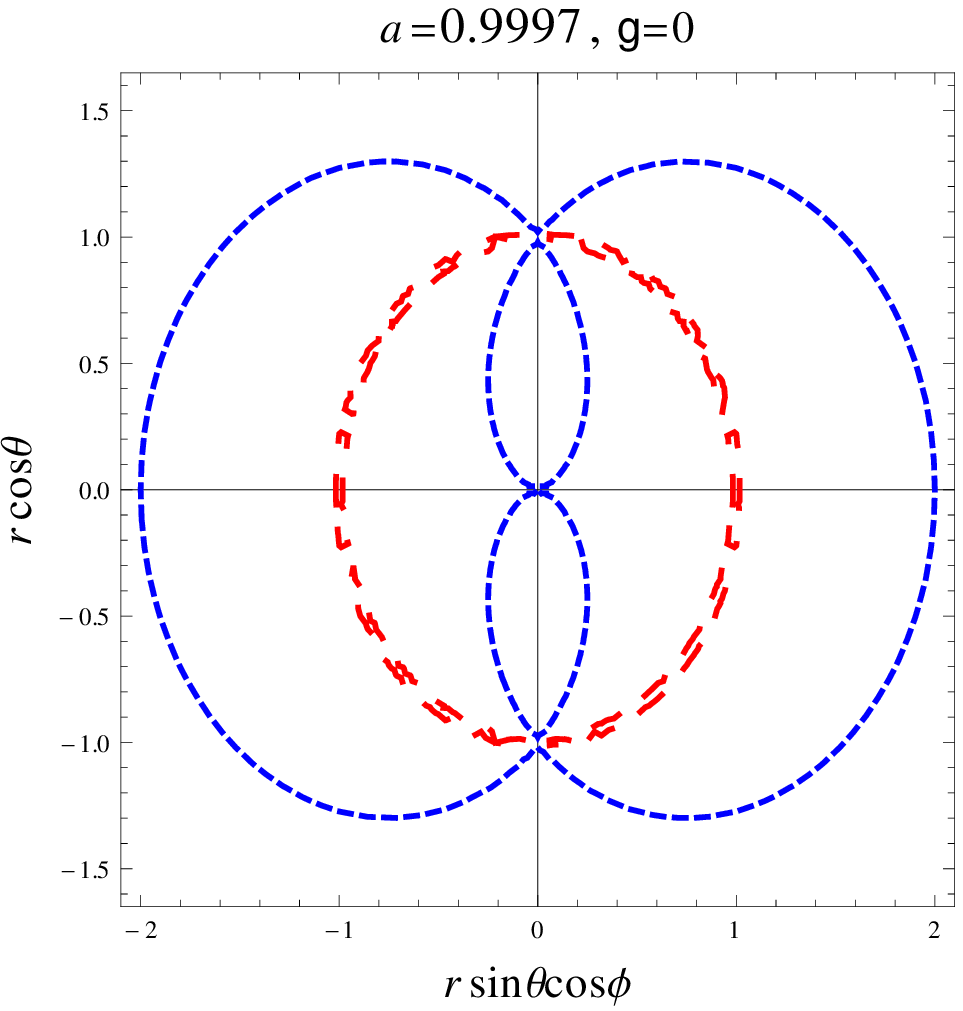}\hspace{-0.2cm}
		\includegraphics[scale=0.42]{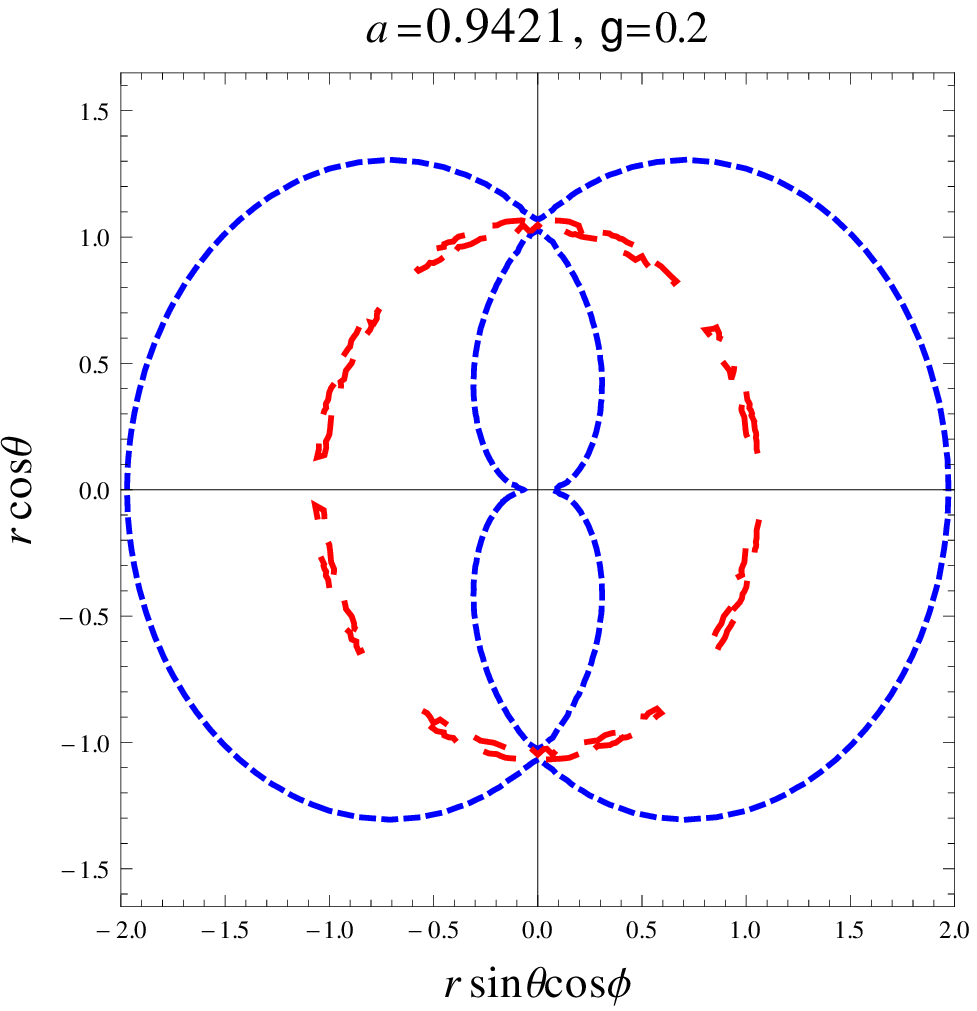}\hspace{-0.2cm}
		\includegraphics[scale=0.42]{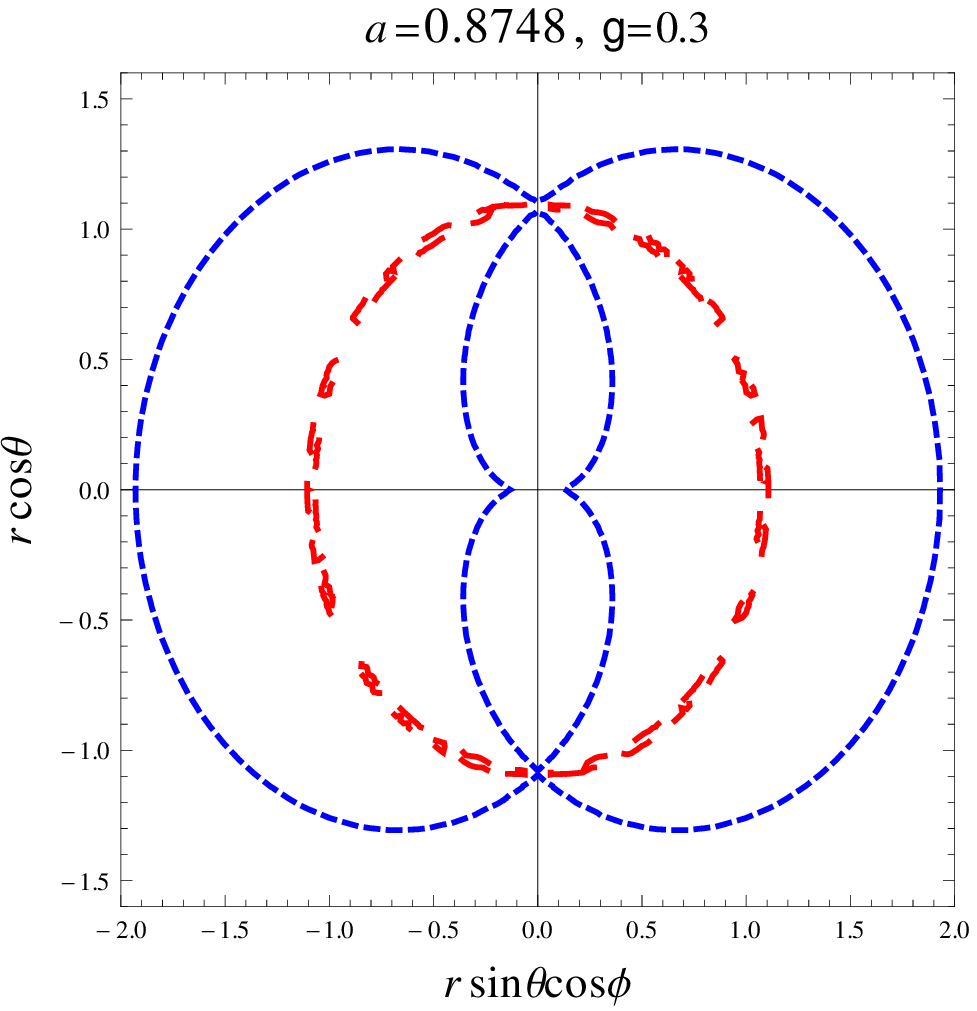}\hspace{-0.2cm}
	   &\includegraphics[scale=0.42]{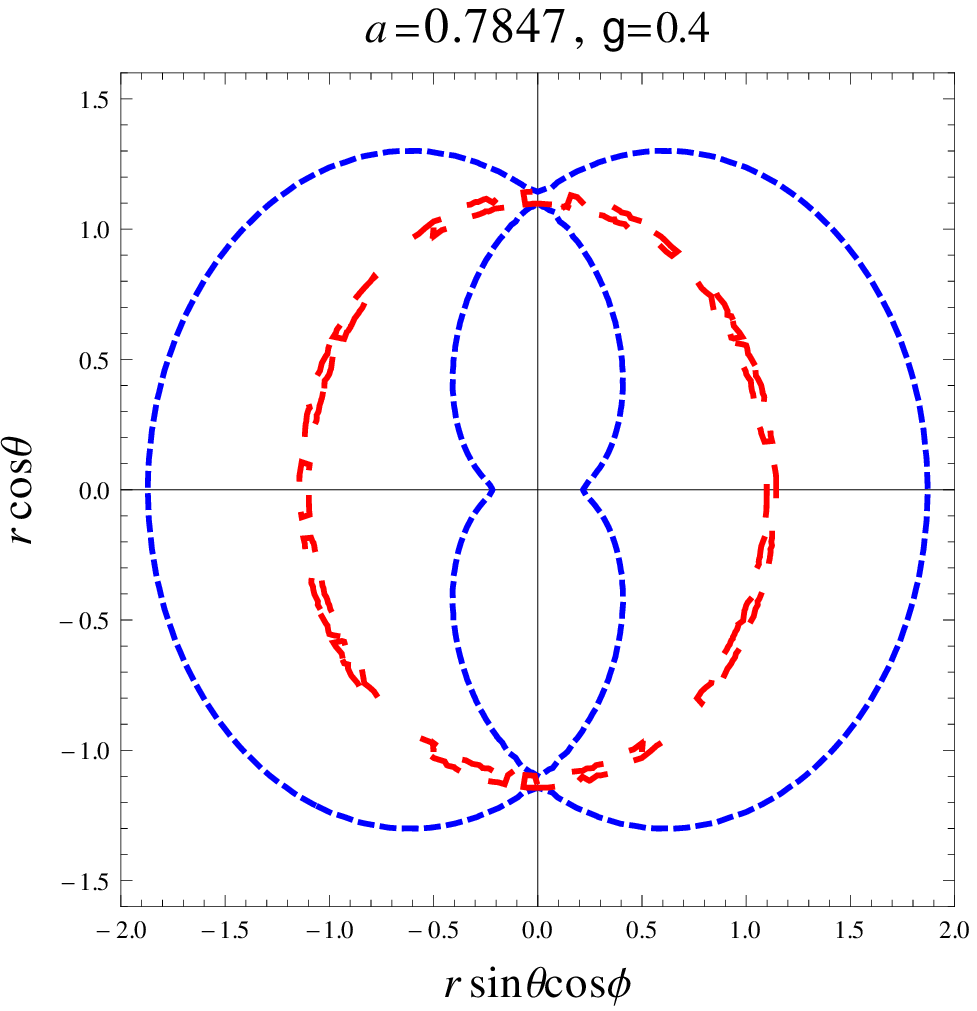}
\end{tabular}
	\caption{Plot showing the variation of the shape of ergosphere for $a \approx a_E$ (extremal black hole) in $xz$-plane with parameter $g$, for different values of $a$, of the rotating Bardeen regular black hole.}\label{fig6}
\end{figure*}

\section{rotating Bardeen regular black hole}
\label{bh}
We do not have a definite quantum theory of gravity, hence important direction of research is to consider regular rotating models to solve singularity problem of standard Kerr black hole. These are called non-Kerr black hole with different spin. Bambi \cite{Bambi:2013ufa}, starting from regular Bardeen metric via Newman-Janis algorithm \cite{Newman:1965tw} constructed a Kerr-like regular black hole solution, which in the Boyer-Lindiquist coordinates reads
\begin{eqnarray}\label{bar}
d{s}^2 &=& -\left(1-\frac{2mr}{\Sigma}\right)dt^2 -\frac{4amr \sin^2 \theta}{\Sigma}dtd\phi +\frac{\Sigma}{\Delta}dr^2
\nonumber \\
&+& \Sigma d\theta^2 +\left(r^2+a^2+\frac{2a^2mr \sin^2 \theta}{\Sigma}\right)\sin^2 \theta d\phi^2,
\end{eqnarray}
where
\begin{eqnarray}
\Sigma =r^2 + a^2\cos^2\theta,\;\;\;\;\; \Delta = r^2-2mr+a^2,
\end{eqnarray}
and
\begin{eqnarray}\label{m}
m\rightarrow m_{\zeta, \lambda}(r,\theta) = M\left(\frac{r^{2+\zeta} \Sigma^{-\zeta/2}}{r^{2+\zeta} \Sigma^{-\zeta/2}+g^2r^{\lambda} \Sigma^{-\lambda/2}}\right)^{3/2},
\end{eqnarray}
where $m=m_{\zeta, \lambda}(r,\theta)$ is a function of both $r$ and $\theta$, $\zeta$, $\lambda$ are two real numbers, and $a$ is rotation parameter. The parameter $g$ is the magnetic charge of non-linear electrodynamics which measures deviation from the Kerr black hole, when we switch off non-electrodynamics ($g=0$), one recovers Kerr metric. It turns out that curvature invariants are regular everywhere, when $g\neq 0$, including at origin \cite{Bambi:2013ufa}.  For definiteness, throughout this paper, we shall call the metric (\ref{bar}) as rotating Bardeen regular black hole.

The Boyer-Linquist coordinates are most widely used by astrophysicists as in these coordinates, the rotating black hole has just one off diagonal term. The rotating Bardeen, like the Kerr metric, in the Boyer-Lindquist coordinates does not depend on $t, \phi$, which means the Killing symmetries, and the Killing vectors are given by $\eta^{\mu} = \delta_t^{\mu}$ and $\xi^{\mu} = \delta_{\phi}^{\mu}$, with $\delta_{a}^{\mu}$ the Kronecker delta. The existence of the two Killing vectors $\eta^{\mu}$  and $\xi^{\mu}$ implies that the corresponding momenta of a test particle, $p_t$ and $p_{\phi}$, are constants of the motion.
\begin{widetext}

\begin{table}
    \caption{The values of horizons of rotating Bardeen regular black hole with parameters $M=1$ and  $\theta=\pi/2$ ($\delta_{h}^{g}=r^{+}_{H}-r^{-}_{H}$).}\label{table1}
  \resizebox{\textwidth}{!}{  
        \begin{tabular}{| c| c c c | c c  c | c c c | c c c |}
            \hline
   &\multicolumn{3}{c}{$g=0$}  &\multicolumn{3}{c}{$g=0.2$}  &\multicolumn{3}{c}{$g=0.3$}  &\multicolumn{3}{c}{$g=0.4$}\\
            \hline
$a$  & $r^{-}_{H}$ & $r^{+}_{H}$ & $\delta_{h}^{g}$ & $r^{-}_{H}$ & $r^{+}_{H}$ & $\delta_{h}^{g}$ & $r^{-}_{H}$ & $r^{+}_{H}$ & $\delta_{h}^{g}$ & $r^{-}_{H}$ & $r^{+}_{H}$ & $\delta_{h}^{g}$  \\
            \hline
0.5  & 0.13397 & 1.86603  & 1.73206 & 0.29652 & 1.82784  & 1.53132 & 0.40030 & 1.77671 & 1.37641 & 0.52235 & 1.69683  & 1.17448 \\
0.6  & 0.20000 & 1.80000  & 1.60000 & 0.36555 & 1.75683  & 1.39128 & 0.48388 & 1.69783 & 1.21395 & 0.62877 & 1.60175  & 0.97298\\
0.7  & 0.28585 & 1.71414  & 1.45829 & 0.45416 & 1.66265  & 1.20849 & 0.59211 & 1.58940 & 0.99729 & 0.77907 & 1.45731  & 0.67824 \\
0.8  & 0.40000 & 1.60000  & 1.20000 & 0.57655 & 1.53228  & 0.95573 & 0.75247 & 1.42484 & 0.67237 & $-$     & $-$  & $-$\\
$a_{E}^*$ & 1.00000  & 1.00000 & 0.00000 & 1.04769  & 1.04769 & 0.00000 & 1.08604 & 1.08604 & 0.00000  & 1.11879 & 1.11879 & 0.00000\\
            \hline 
        \end{tabular}
}
 \resizebox{\textwidth}{!}{  
 {$a_{E}^*=1, 0.9488364581472$, $0.875075019710$, and $0.785157362359$, which respectively correspond to  $g=0, 0.2$, $0.3$, and $0.4$.}}
\end{table}

\begin{table}
       \caption{The values of outer static limit surface and event horizon of rotating Bardeen regular black hole with parameters $\zeta=2$, $\lambda=3$, $M=1$ and  $\theta=\pi/4$ ($\delta_{e}^{g}=r^{+}_{sls}-r^{+}_{H}$). }\label{table2}
      \resizebox{\textwidth}{!}{
        \begin{tabular}{| c | c c c| c c c | c c c | c c c |}
            \hline
            &\multicolumn{3}{c}{$g=0$}  &\multicolumn{3}{c}{$g=0.2$}  &\multicolumn{3}{c}{$g=0.3$} &  \multicolumn{3}{c}{$g=0.4$}\\
            \hline
$a$ & $r^{+}_{H}$ & $r^{+}_{sls}$ & $\delta_{e}^{g}$  & $r^{+}_{H}$ & $r^{+}_{sls}$ & $\delta_{e}^{g}$ & $r^{+}_{H}$ & $r^{+}_{sls}$  & $\delta_{e}^{g}$ & $r^{+}_{H}$ & $r^{+}_{sls}$  & $\delta_{e}^{g}$ \\
            \hline
 0.5   & 1.86603  & 1.93541  & 0.06938  & 1.82855  & 1.90211  & 0.07356 & 1.77855  & 1.85831  & 0.07976 & 1.70092 & 1.79209     & 0.09117 \\
 0.6   & 1.80000  & 1.90554  & 0.10554  & 1.75807  & 1.87081  & 0.11274 & 1.70115  & 1.82492  & 0.12377 & 1.60973 & 1.75498     & 0.14525\\
 0.7   & 1.71414  & 1.86891  & 0.15477  & 1.66490  & 1.83227  & 0.16737 & 1.59578  & 1.78355  & 0.18777 & 1.47589 & 1.70843     & 0.23254\\
 0.8   & 1.60000  & 1.82462  & 0.22462  & 1.53680  & 1.78541  & 0.24861 & 1.44038  & 1.73279  & 0.29241 & 1.19504 & 1.65023     & 0.45519\\
 0.9    & 1.43589  & 1.77136  & 0.33547  & 1.33282  & 1.72860 & 0.39578  & $-$     & 1.67041    & $-$   & $-$    & 1.57651    & $-$\\	
            \hline
        \end{tabular}
}
\end{table}

\end{widetext}

\subsection{Horizons and ergosphere}
The metric (\ref{bar}), is singular at $\Delta=0$, which corresponds to event horizon of a black hole. The horizons of the rotating Bardeen regular black hole are solution of
\begin{equation}\label{ehc}
\left(\eta^{\nu}\xi_{\mu}\right)^2 - (\eta^{\nu}\eta_{\nu}) (\xi^{\mu}\xi_{\mu}) =0,
\end{equation}
which leads to
\begin{eqnarray}\label{eh2}
(r^2+a^2)^2 (r^{2+\zeta} \Sigma^{-\zeta/2}+g^2r^{\lambda}\Sigma^{-\lambda/2})^3
\nonumber \\
-4M^2r^{8+3\zeta} \Sigma^{-3\zeta/2}=0.
\end{eqnarray}
which depends on $a$, $g$, and $\theta$, and is different from
the Kerr black hole where it is $\theta$ independent.
The numerical analysis of (\ref{eh2}) suggest the possibility of two roots for a set of values of parameters which corresponds to two horizons of rotating Bardeen regular black hole. The behavior of event horizon is shown in Figs.~\ref{fig1},~\ref{fig2} and~\ref{fig3} for different values of $a$ and $g$. The Fig.~\ref{fig3} imply that for, $a<a_{E}$, there exists a set of values of parameters for which one gets two horizons and when $a=a_{E}$, the two horizons coincide, i.e., extremal black hole with degenerate horizons (cf. Table~\ref{table1}). An extremal  black holes occur when $\Delta=0$ has a double root, i.e., when the two horizons coincides. When $\Delta=0$ has no root, i.e., no horizon exists which mean there is no black hole (cf. Figs.~\ref{fig1} and \ref{fig2}). There exists an upper bound on the spin parameter $a^E{*} \leq 0.998$ for astrophysical black holes, which is called Thorne's bound \cite{thorne}. In the equatorial plane ($\theta=\pi/2$), the mass function $ m_{\zeta, \lambda}(r,\theta) $  does not
depend on the parameters $\zeta$ and $\lambda$, and so is 
Eq.~(\ref{eh2}). This is also the case when both $\zeta = \lambda
=0.$  Thus, the two cases $\theta=\pi/2$, and $\zeta = \lambda =0$
(for any $\theta$), will have same horizon structure.
The non trivial effect of the parameters $\zeta$ and $\lambda$ ($\theta \neq \pi/2$) on the horizon structure is shown in Fig.~\ref{fig1}. Note that the horizon is called extremal when $r= r^{E}_{H}$ is a double root of $\Delta=0$ when $a=a_{E}$. It is see that  for $\zeta=2$, $\lambda=3$, $\theta=\pi/4$ and $g=0.2$, we have $a_{E}=0.95145807$ and $r^{E}_{H}=1.02899$, and for $g=0.4$, one gets $a_{E}=0.80802796$ and $r^{E}_{H}=1.08549$. Thus, the extremal value of the rotation parameter $a_{E}$ decreases due to the presence of the magnetic charge $g$ when compared with the Kerr black hole.  Thus, for each $g$, there exist an extremal rotating Bardeen black hole.

\begin{figure*}
\begin{tabular}{c c}
 \includegraphics[scale=0.66]{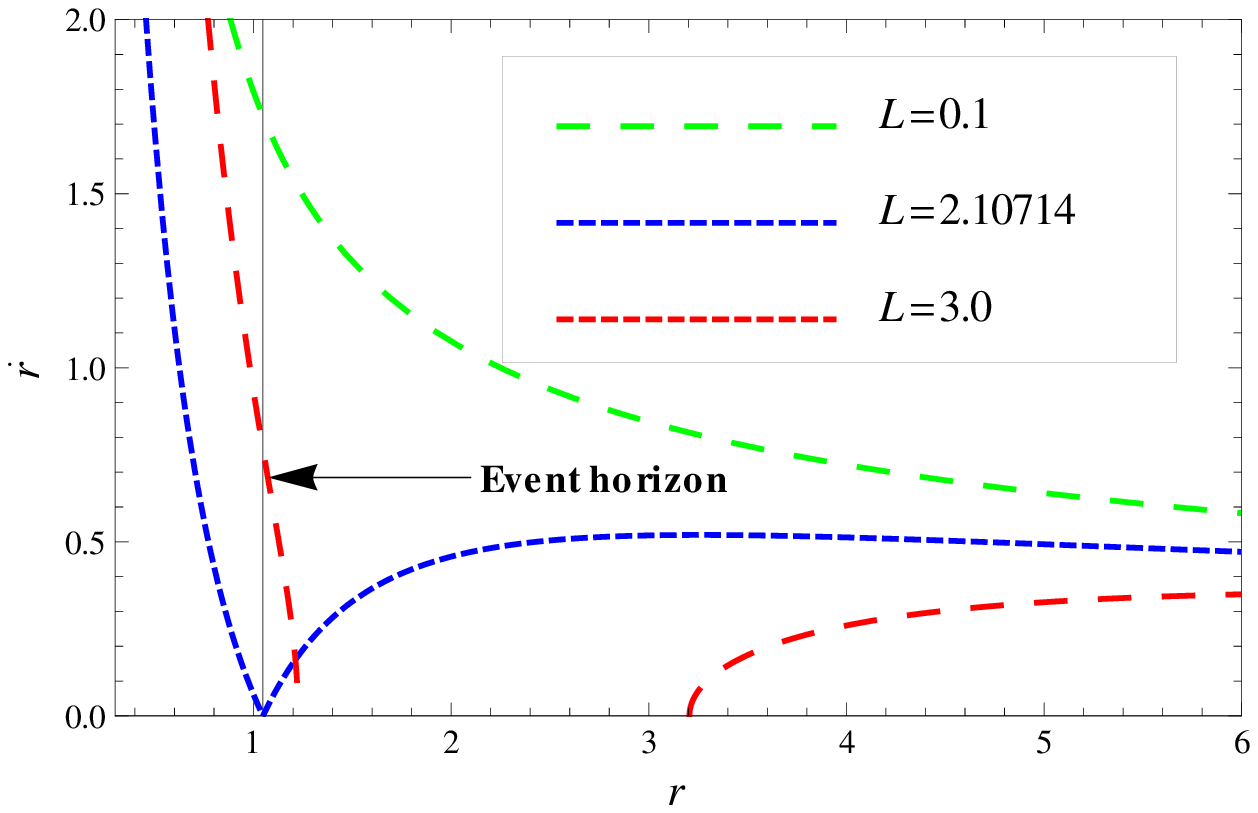}\hspace{-0.9cm}
&\includegraphics[scale=0.66]{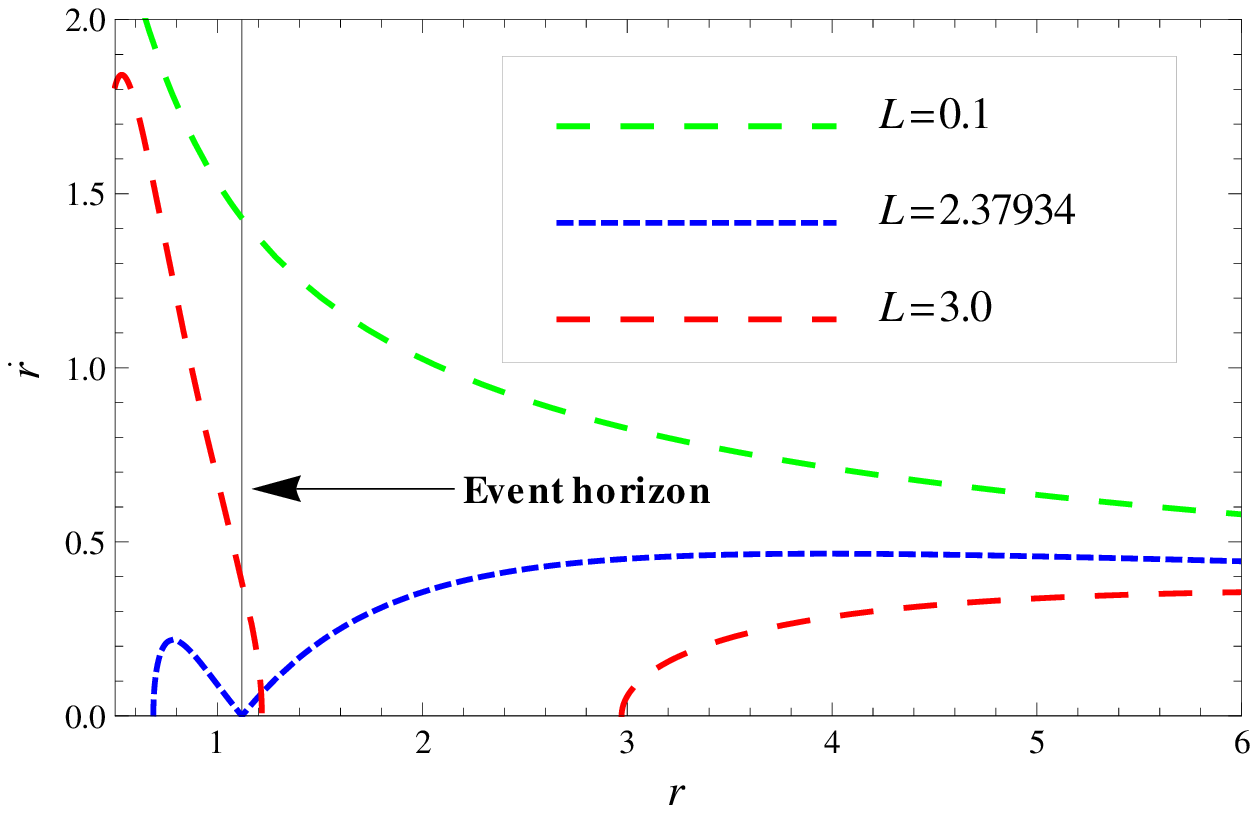}
\end{tabular}
 \caption{The behavior of $\dot{r}$ vs $r$ for extremal black hole. Panel (a) for magnetic charge $g=0.2$ and $a=0.942439535325$. Panel (b) for magnetic charge $g=0.4$ and $a=0.7851573623591$.}\label{fig7}
\end{figure*}

\begin{figure*}
\begin{tabular}{c c}
 \includegraphics[scale=0.62]{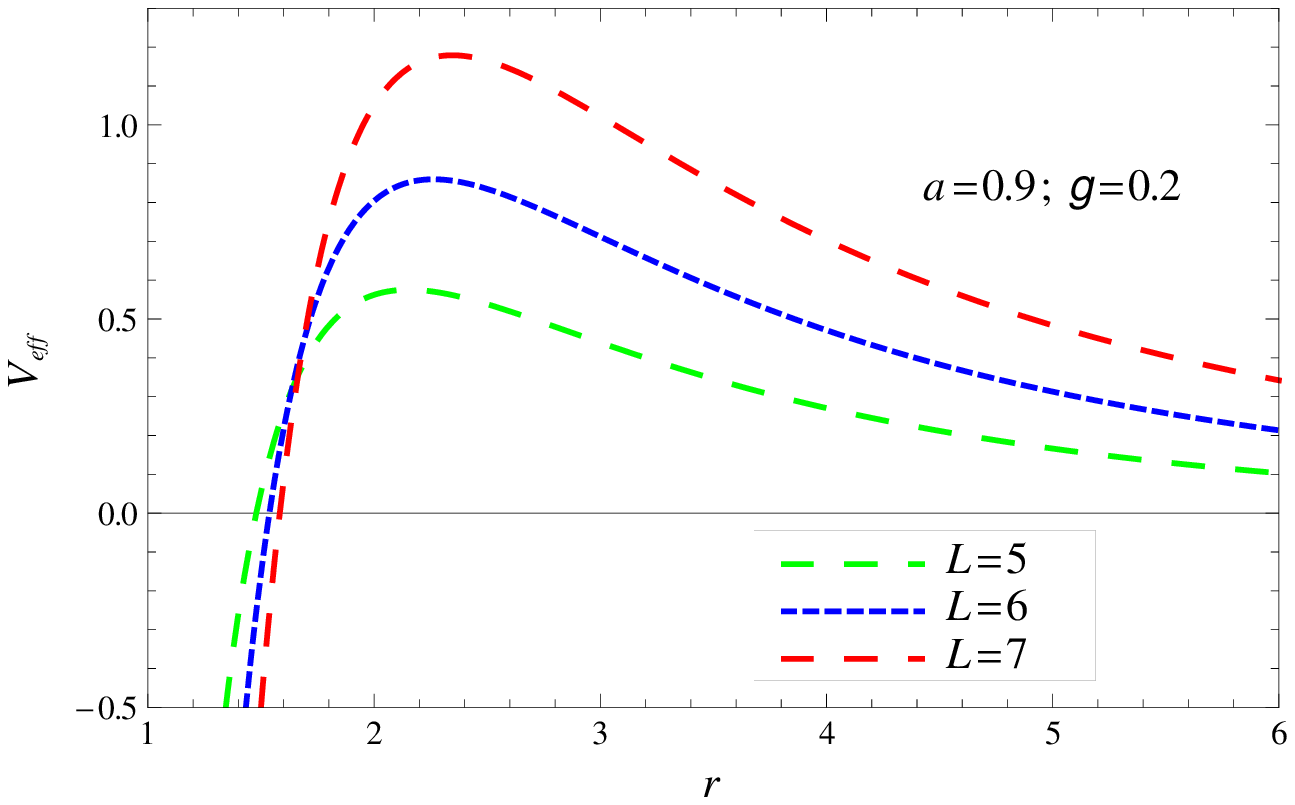}\hspace{-0.2cm}
&\includegraphics[scale=0.62]{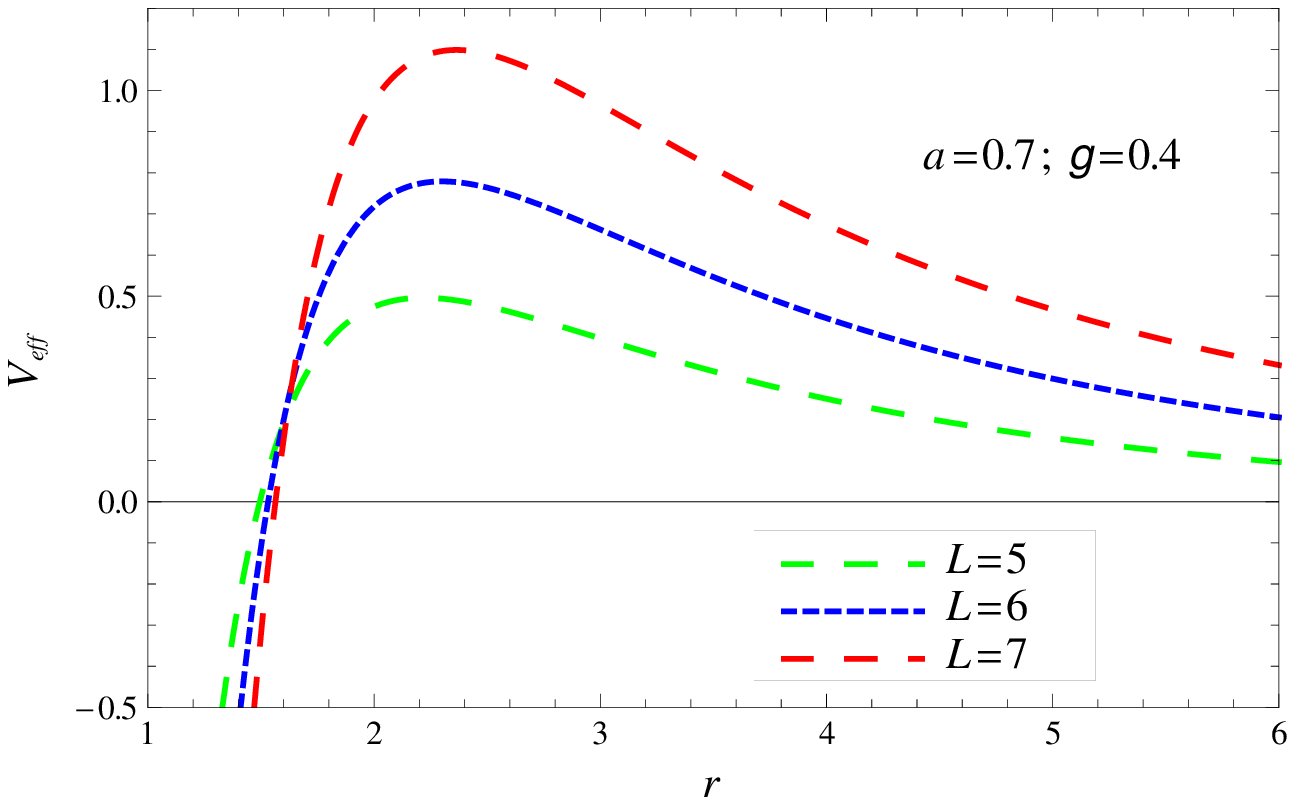}
\end{tabular}
 \caption{Plot showing the behavior of $V_{eff}$ vs $r$ for different angular momentum $L$.}\label{fig8}
\end{figure*}

The static limit surface or infinite redshift surface of a black hole is a surface where the time-translation Killing  vector becomes null, $\eta^{\mu}\;\eta_{\mu}=0$. An infinite red-shift surface is referred as null Killing surfaces to  distinguish them from the null surfaces corresponding to the horizons.  To find the null Killing surfaces, one sets the $g_{tt}$ component of the metric tensor in Eq.~(\ref{bar}) equal to zero. For the rotating Bardeen regular black hole, we find that $g_{tt}=0$ gives,
\begin{eqnarray}\label{eh}
\Sigma ^2 (r^{2+\zeta} \Sigma^{-\zeta/2}+g^2r^{\lambda}\Sigma^{-\lambda/2})^3
\nonumber \\
-4M^2r^{8+3\zeta} \Sigma^{-3\zeta/2}=0.
\end{eqnarray}
In the Fig.~\ref{fig4}, we depict the possible roots of equation $g_{tt}=0$ with different combinations of the parameters $a$ and $g$ and different values of $\theta$. The event horizon of the rotating Bardeen regular metric (\ref{bar}) is located at $r=r_{H}^{+}$, the larger root of $\Delta=0$, and the static limit surface is located at $r=r_{sls}^{+}$ which is zero of $g_{tt}=0$. Observing the outer event horizon and stationary
limit surface of the rotating Bardeen regular black hole, it is verified that the stationary limit surface
always lies outside the event horizon for all values of $g$.
Hence, as in the Kerr black hole, we call the region between two surfaces as the ergosphere, which lies outside of a black hole. The ergosphere is given by $r_{H}^{+}<r<r_{sls}^{+}$, and important feature of the ergosphere  is that the timelike killing vector becomes spacelike after crossing the static limit surface, i.e., in the ergosphere $\eta^{\mu}\;\eta_{\mu}=g_{tt}>0$. The existence of the ergosphere allows various kinds of energy extraction mechanisms for a rotating black hole. It has been suggested that ergosphere can be used to extract energy from rotating black holes through the Penrose process \cite{Penrose}. Further, observer moving along the time-like geodesics is always stationary in the ergosphere due to the "framedragging effect" \cite{Chandra}. Its shape is that of an oblate spheroid - bulging at the equator, and flattened at the poles of the rotating black hole.  We have studied how the parameters $a,g$ affect the shape of the ergosphere, and the behavior of ergosphere, for $a<a_E$ is depicted in Figs.~\ref{fig5} and for $a \approx a_E$ in ~\ref{fig6}. It can be seen that  ergosphere is sensitive to the parameter $g$, the ergosphere area enlarges with increase in the parameter $g$, as well as with $a$ (cf. Table~\ref{table2}). In the extremal case $a \approx a_E$ (Fig.~\ref{fig6}), the inner and outer horizons that coincide, and the thickness of the ergosphere still increases with increase in the when the parameter $g$.

\section{Equations of motion and the effective potential }
\label{eqms}
In this section, we would like to study the equations of motion of a particle with rest mass $m_{0}$ falling in the background of the rotating Bardeen regular black hole. Henceforth, we shall restrict our discussion to the case of equatorial plane ($\theta=\pi/2 $), which simplifies the mass function (\ref{m}) to
\begin{equation}\label{m1}
m= M\left(\frac{r^2}{r^2+g^2}\right)^{3/2}.
\end{equation}
It is easy to see that the mass function (\ref{m1}) can be also obtained from (\ref{m}) for $\zeta=\lambda=0$.
There are two killing vectors, timelike ($ \eta_{t}^{\mu} $) and axial ($ \xi_{\phi}^{\mu} $) killing vectors. So there must be two conserved quantities corresponding to these killing vectors. The conserved quantities corresponds to these timelike and axial killing vectors are energy ($E$) and angular momentum ($L$) respectively. The conserved quantities at equatorial plane are defined by the following equations
\begin{equation}\label{kt}
-E = g_{\mu \nu} \eta_{t}^{\mu}u^{\nu} = g_{tt}u^{t} +  g_{t \phi}u^{\phi},
\end{equation}
\begin{equation}\label{kp}
L = g_{\mu \nu} \xi_{\phi}^{\mu}u^{\nu} = g_{\phi t}u^{t} +  g_{\phi \phi}u^{\phi},
\end{equation}
where $u^{\nu}$ is the four-velocity of the particle. To calculate the $E_{CM}$ for the colliding particles, first of all we need to calculate the four-velocities of the particles. The four-velocities are calculated by solving the Eq.~(\ref{kt}) and Eq.~(\ref{kp}) simultaneously and using the condition $u_{\nu}u^{\nu}=-m_{0}^2$. Hence the four-velocities of the falling particles have the following form
\begin{equation}\label{eqm1}
 u^{t} = \frac{1}{r^2} \Big[-a (aE - L) + \left(r^2 + a^2\right) \frac{T}{\Delta}\Big],
\end{equation}
\begin{equation}\label{eqm2}
 u^{ \phi} = \frac{1}{r^2} \Big[-(aE - L ) + \frac{a T}{\Delta}\Big],
\end{equation}
\begin{equation}\label{eqm3}
 u^{r} = \pm \frac{1}{r^2} \sqrt{T^2 -\Delta \left[m_0^2 r^2  + (L-a E)^2\right]},
\end{equation}
where $ T = E (r^2 + a^2) -La $. In Eq.~(\ref{eqm3}) the $+$ sign corresponds to the outgoing geodesic and $-$ sign corresponds to the incoming geodesics. The effective potential is calculated as
\begin{equation}
\frac{1}{2} (u^{r})^2+ V_{eff}=0.
\end{equation}
Hence, the form of effective potential is looks like as
\begin{equation}
 V_{eff} =  -\frac{[E(r^2+a^2)-La]^2 -\Delta [m_0^2 r^2  + (L-a E)^2]}{2r^4}.
\end{equation}
The range of the angular momentum for the falling particles are calculated by the following equations
\begin{equation}\label{lim}
V_{eff}=0 \;\;\;\text{and}\;\;\; \frac{dV_{eff}}{dr}=0.
\end{equation}
The plots of $\dot{r}$ vs $r$ can be seen from Fig.~\ref{fig7} for different values of $L$, $a$ and $g$. We can see from this Figure that if the angular momentum of the particle is larger $L>L_{c}$, then the geodesics never fall into the black hole. On the other hand if the angular momentum $L<L_{c}$, then the geodesics always fall into the black hole and if $L=L_{c}$, then the geodesics fall into the black hole exactly at the event horizon. The behavior of effective potential ($V_{eff}$) with radius ($r$) can be seen from Fig~\ref{fig8}.

\begin{table}
\begin{center}
\caption{The limiting values of angular momentum for different extremal cases of rotating Bardeen regular black hole.}\label{table3}
\begin{tabular}{| c c | c | c | c | c |}
 \hline
& $g$     & $a_{E}$          & $r^{E}_{H}$   & $L_{2}$    & $L_{1}$ \\
\hline
 & 0      & 1.0              & 1.00000  		&-4.82843    & 2.00000  \\
 & 0.2    & 0.942439535325   & 1.04769  		&-4.77988    & 2.10714  \\
 & 0.3    & 0.875075019710   & 1.08604  		&-4.72090    & 2.22294  \\
 & 0.4    & 0.785157362359   & 1.11879   	&-4.63854    & 2.37934  \\
 & 0.5    & 0.671964660485   & 1.13979  		&-4.52854    & 2.60418  \\
 & 0.6    & 0.529498982374   & 1.14404  		&-4.38025    & 3.00134  \\
 \hline
\end{tabular}
\end{center}
\end{table}
\begin{table}
\begin{center}
\caption{The limiting values of angular momentum for different non-extremal cases of rotating Bardeen regular black hole.}\label{table4}
\begin{tabular}{| c c | c | c | c | c | c |}
 \hline
& $g$   & $a$ & $r^{+}_{H}$  & $r^{-}_{H}$  & $L_{4}$    & $L_{3}$ \\
\hline
 & 0     & 0.95   & 1.31225  	& 0.68775		&-4.79285    & 2.44721  \\
 & 0.2   & 0.90   & 1.31866  	& 0.78085		&-4.74908    & 2.49499  \\
 & 0.3   & 0.80   & 1.42484  	& 0.75247		&-4.66471    & 2.70926  \\
 & 0.4   & 0.70   & 1.45731  	& 0.77907		&-4.57216    & 2.84820  \\
 & 0.5   & 0.60   & 1.43116  	& 0.83941		&-4.46958    & 2.94361  \\
 & 0.6   & 0.50   & 1.31840  	& 0.96243		&-4.38474    & 3.18263  \\
 \hline
\end{tabular}
\end{center}
\end{table}
For the timelike particles, $u^{t} \geq 0$, then from Eq.~(\ref{eqm1}) the condition
\begin{eqnarray}
 E [r^3+2ma^2+a^2r] \;\;\geq 2mLa,
\end{eqnarray}
must be satisfied, as $r \rightarrow r_{H}^{E}$, this condition reduces to
\begin{equation}
E \geq \frac{a}{2mr^{E}_{H}}L = \frac{a}{a^2+(r^{E}_{H})^2}L=\Omega_{H}L,
\end{equation}
where $ \Omega_{H} $ is angular velocity at event horizon.

\begin{figure*}
\begin{tabular}{c c}
 \includegraphics[scale=0.62]{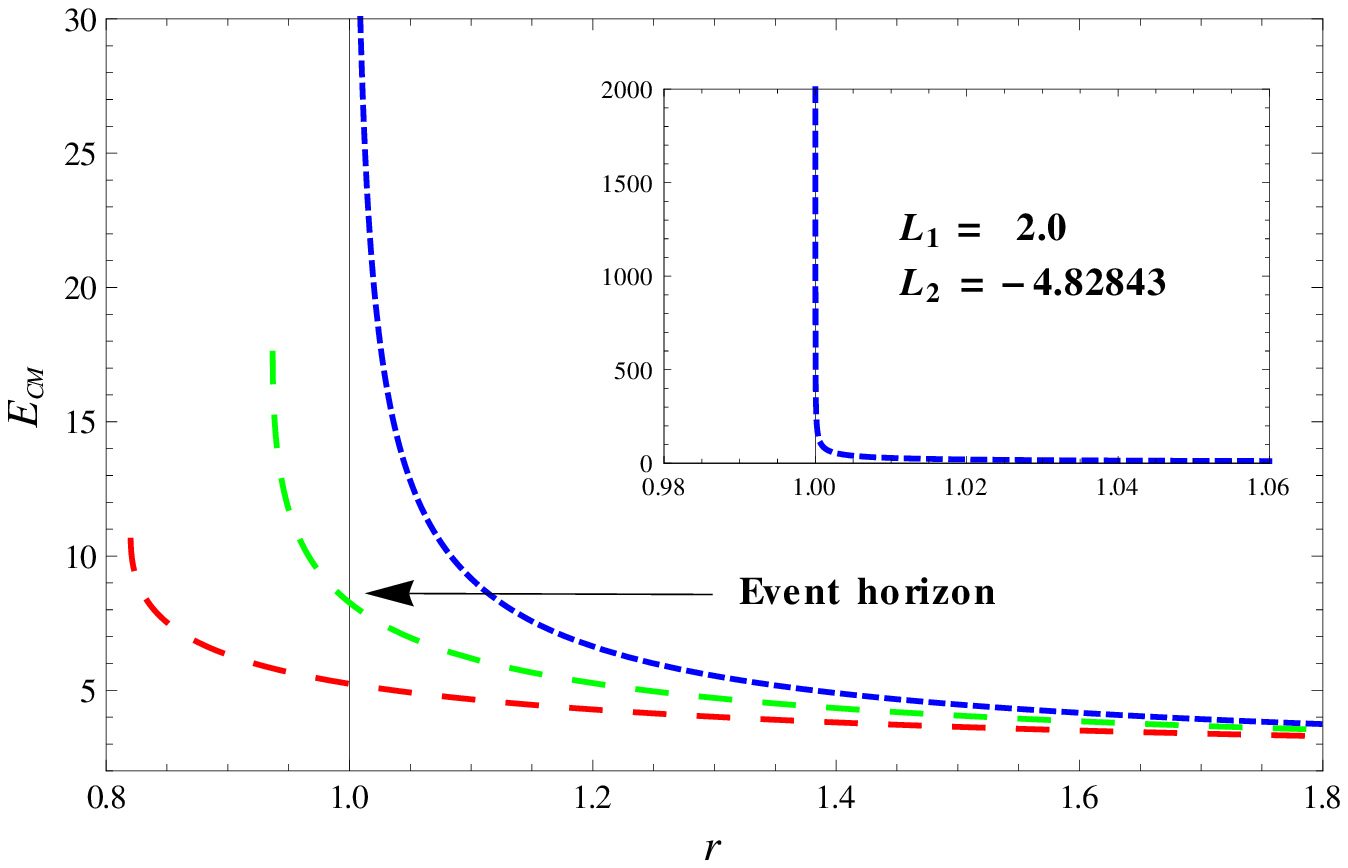}\hspace{-0.2cm}
&\includegraphics[scale=0.62]{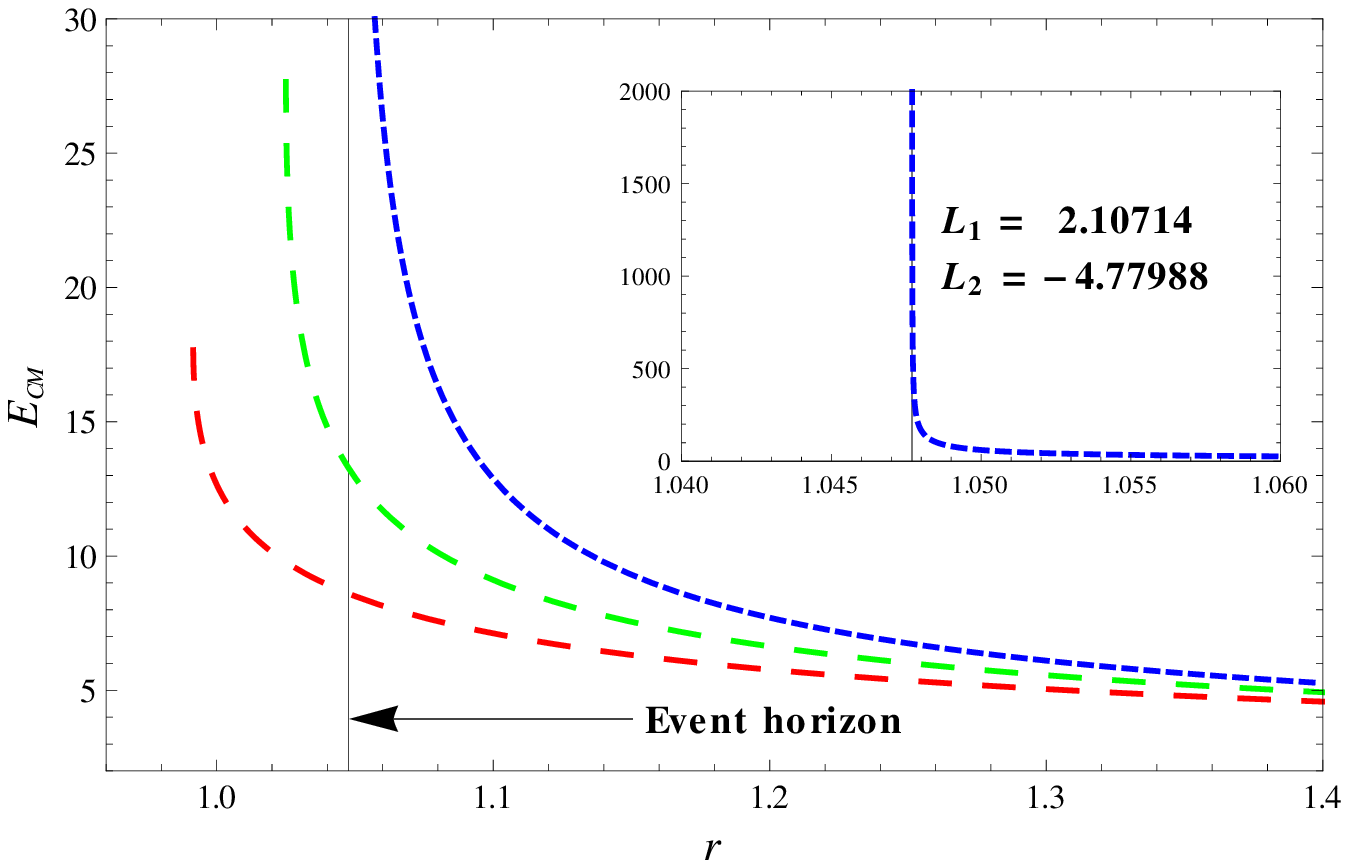}\\
 \includegraphics[scale=0.62]{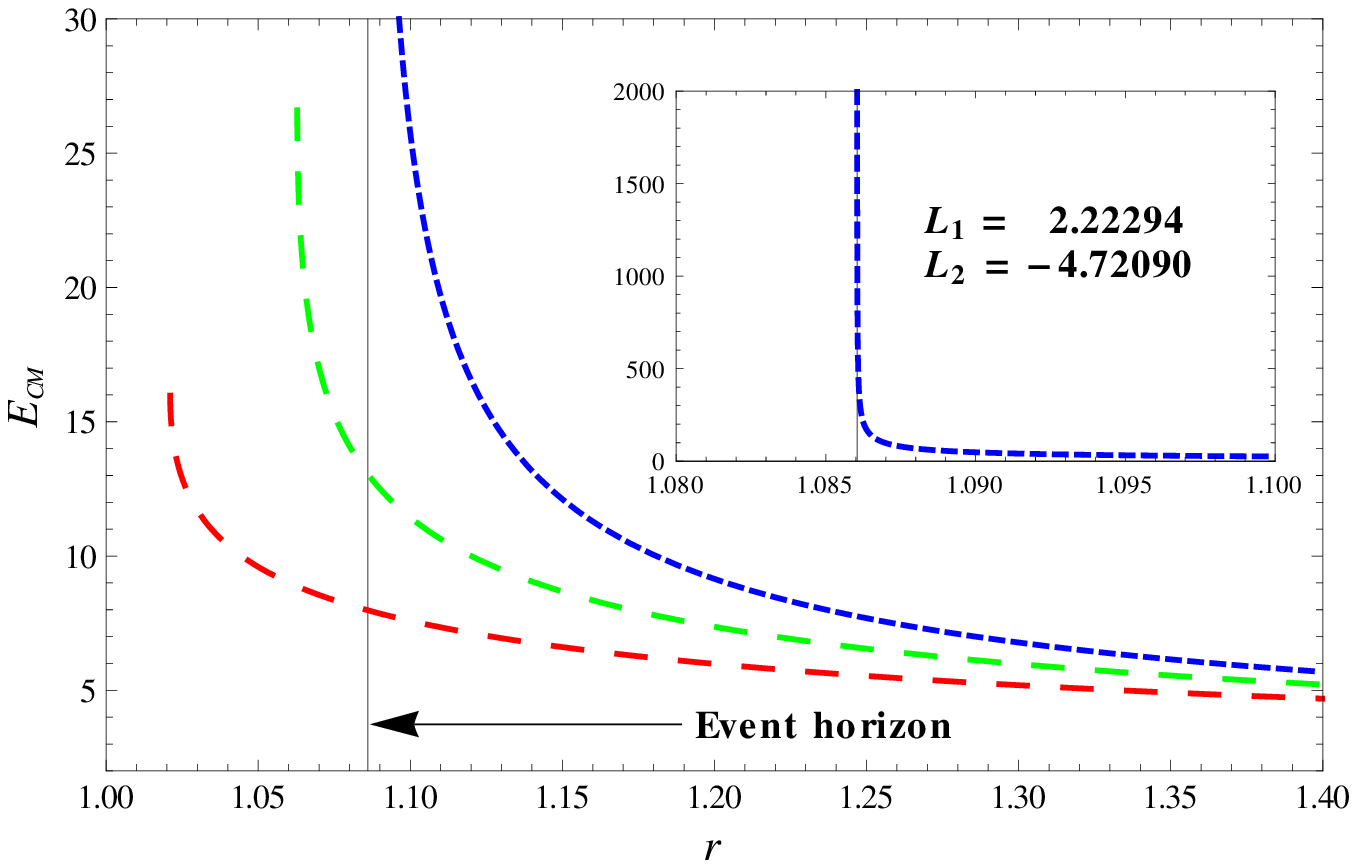}\hspace{-0.2cm}
&\includegraphics[scale=0.62]{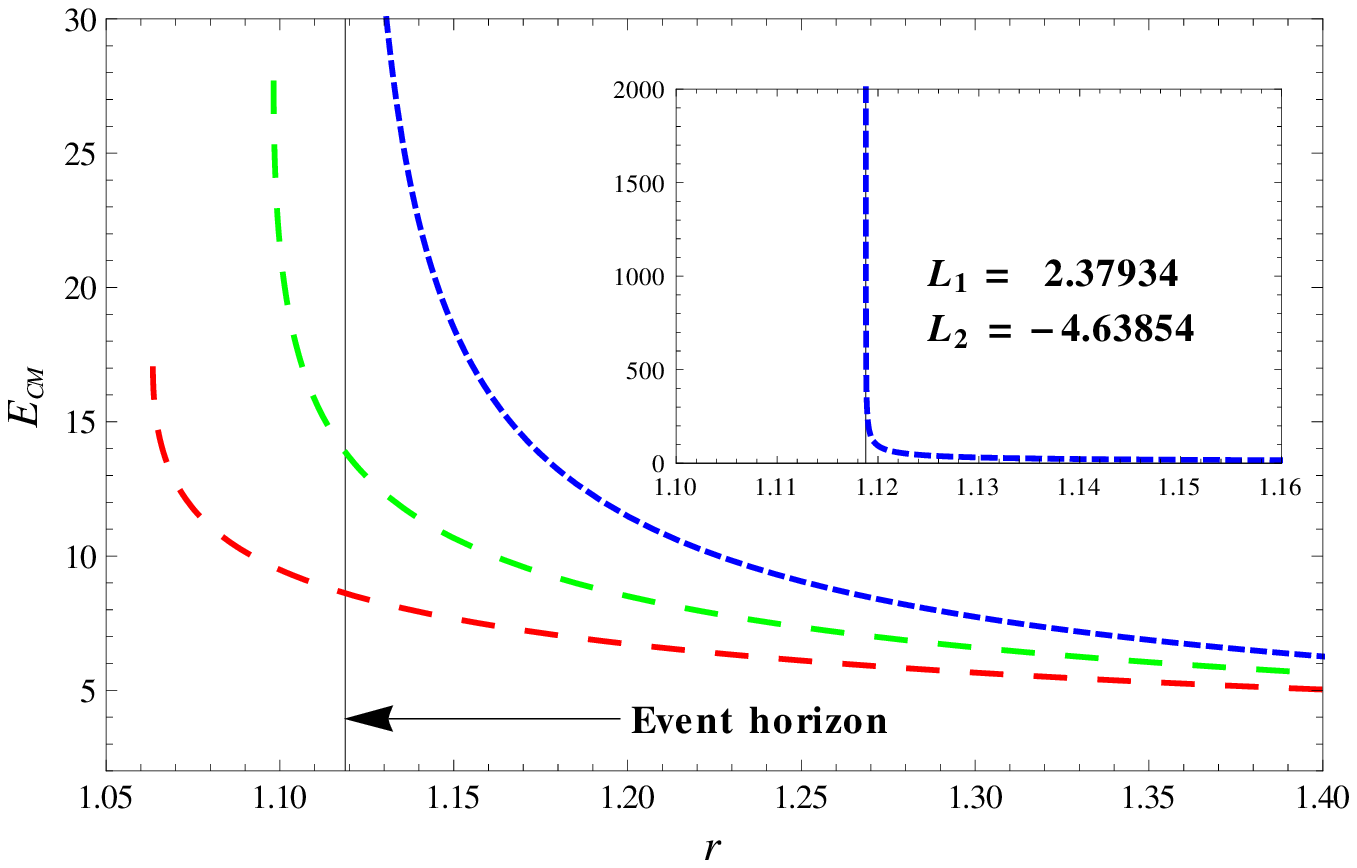}
\end{tabular}
 \caption{Plot showing the behavior of $E_{CM}$ vs $r$ for extremal black hole. (Top) For magnetic charge $g=0$, spin $a_{E}=1$, angular momentum $L_{1}=2.0$ (blue), 1.8 (green), 1.5 (red), and $L_{2}=-4.82843$ (left). For $g=0.2$, $a_{E}=0.942439535325$,  $L_{1}=2.10714$ (blue), 2.02 (green), 1.90 (red),  and $L_{2}=-4.77988$ (right). (Bottom) For  $g=0.3$, $a_{E}=0.875075019710$, $L_{1}=2.22294$ (blue), 2.12 (green), 1.95 (red), and $L_{2}=-4.72090$ (left). For $g=0.4$, $a_{E}=0.7851573623591$, $L_{1}=2.37934$ (blue), 2.27 (green), 2.10 (red), and $L_{2}=-4.63854$ (right).  }\label{fig9}
\end{figure*}

\begin{figure*}
\begin{tabular}{c c}
 \includegraphics[scale=0.62]{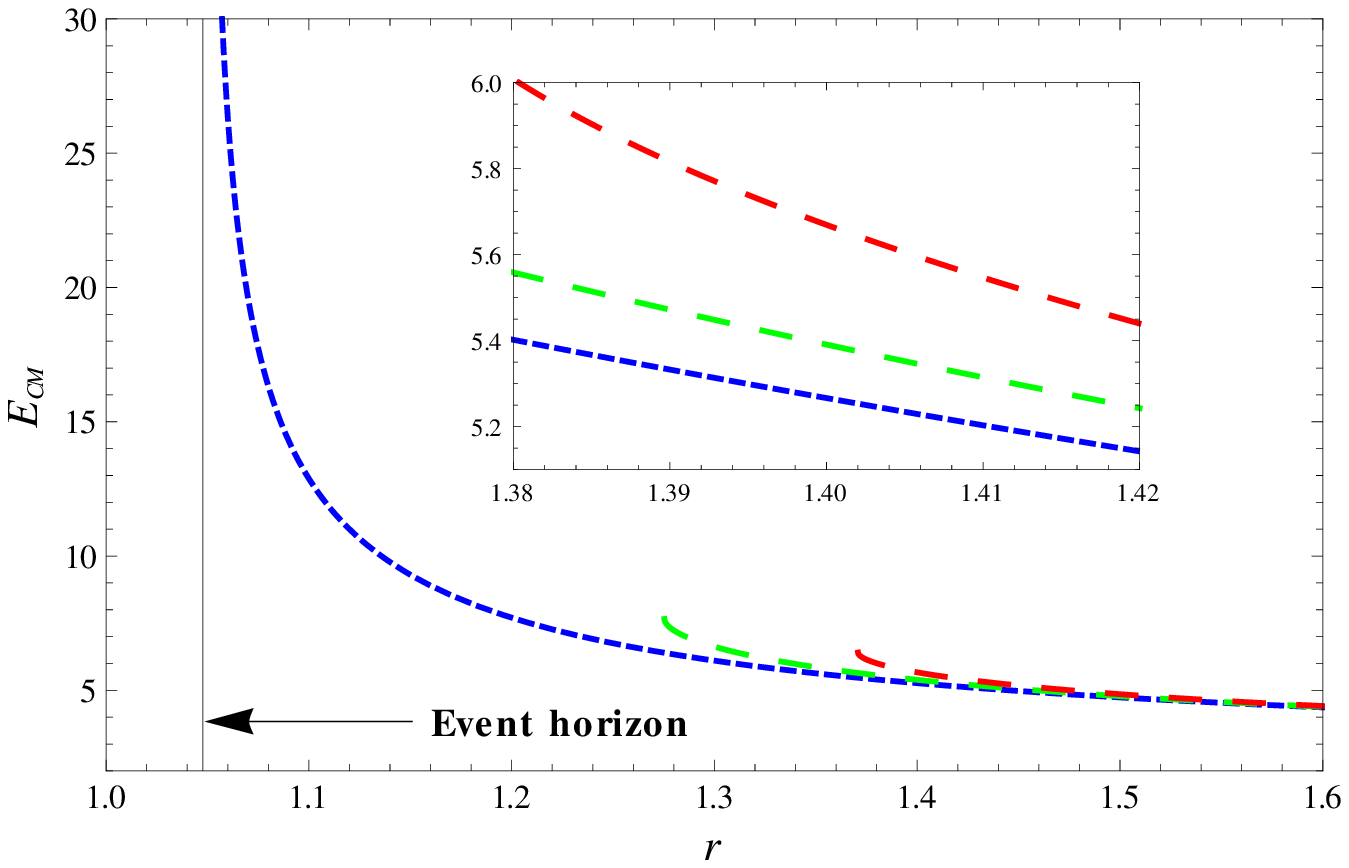}\hspace{-0.2cm}
&\includegraphics[scale=0.62]{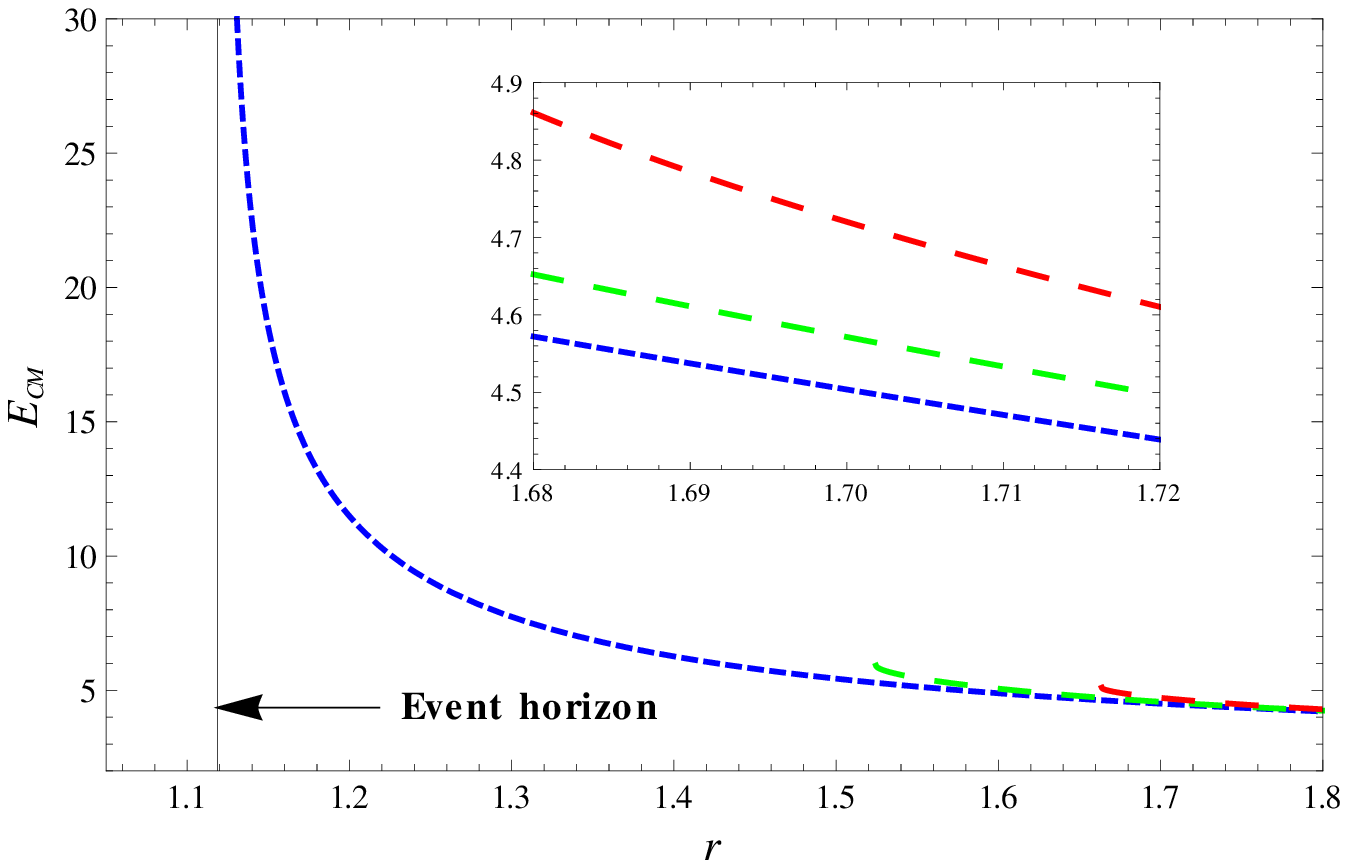}
\end{tabular}
 \caption{Plot showing the behavior of $E_{CM}$ vs $r$. (Left) For $a_{E}=0.942439535325$, $L_{1}=2.10714$, $L_{2}=-4.77988$, and $g=0.2$ (blue), 0.24 (green), 0.28 (red). (Right) For $a=0.7851573623591$, $L_{1}=2.37934$, $L_{2}=-4.63854$, and $g=0.4$ (blue), 0.44 (green), 0.48 (red).}\label{fig10}
\end{figure*}

\begin{figure*}
\begin{tabular}{c c}
 \includegraphics[scale=0.62]{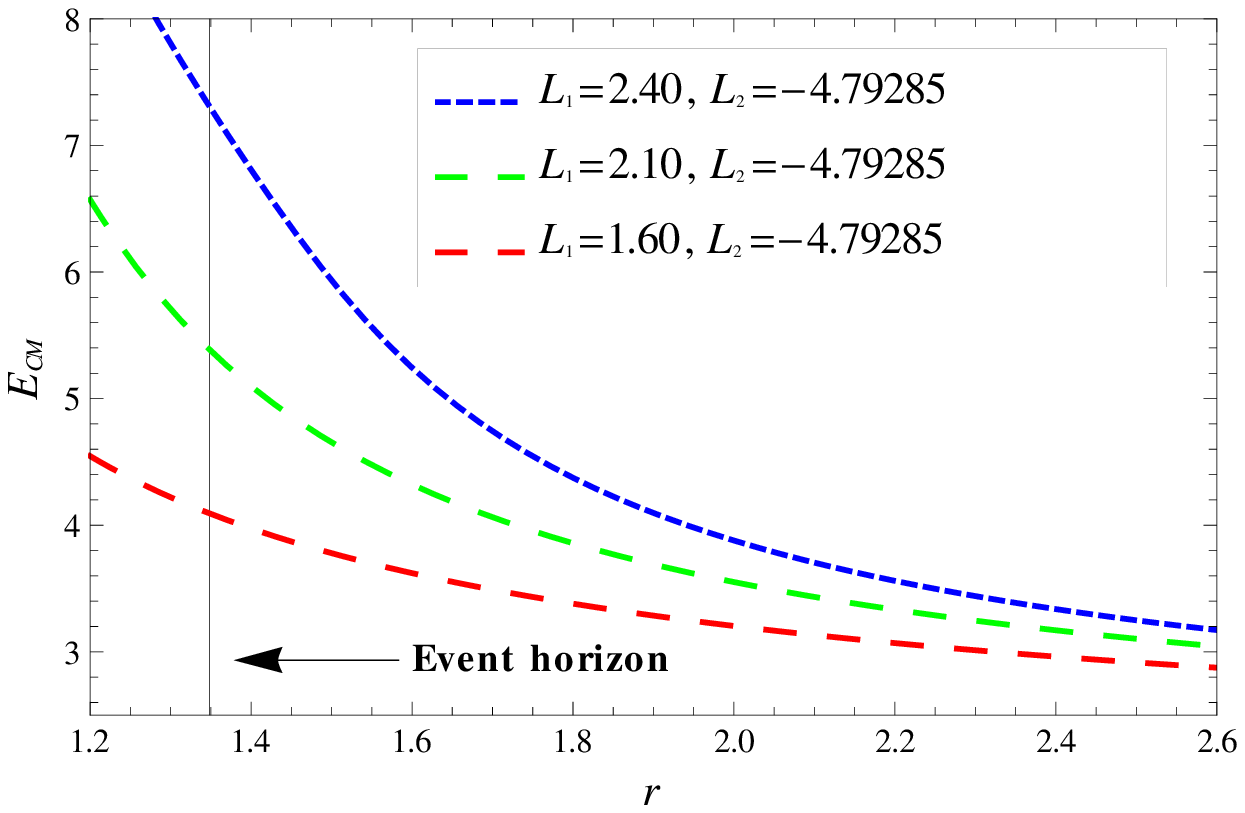}\hspace{-0.2cm}
&\includegraphics[scale=0.62]{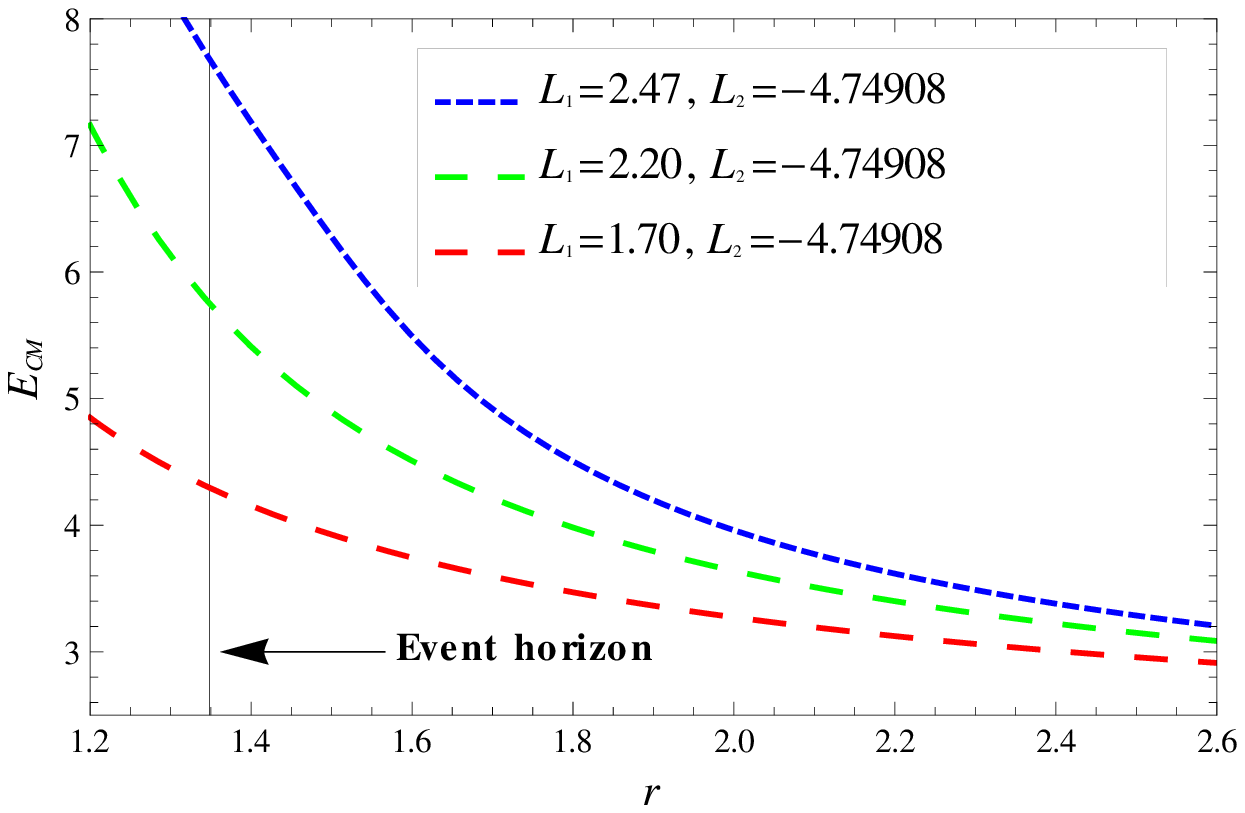}\\
 \includegraphics[scale=0.62]{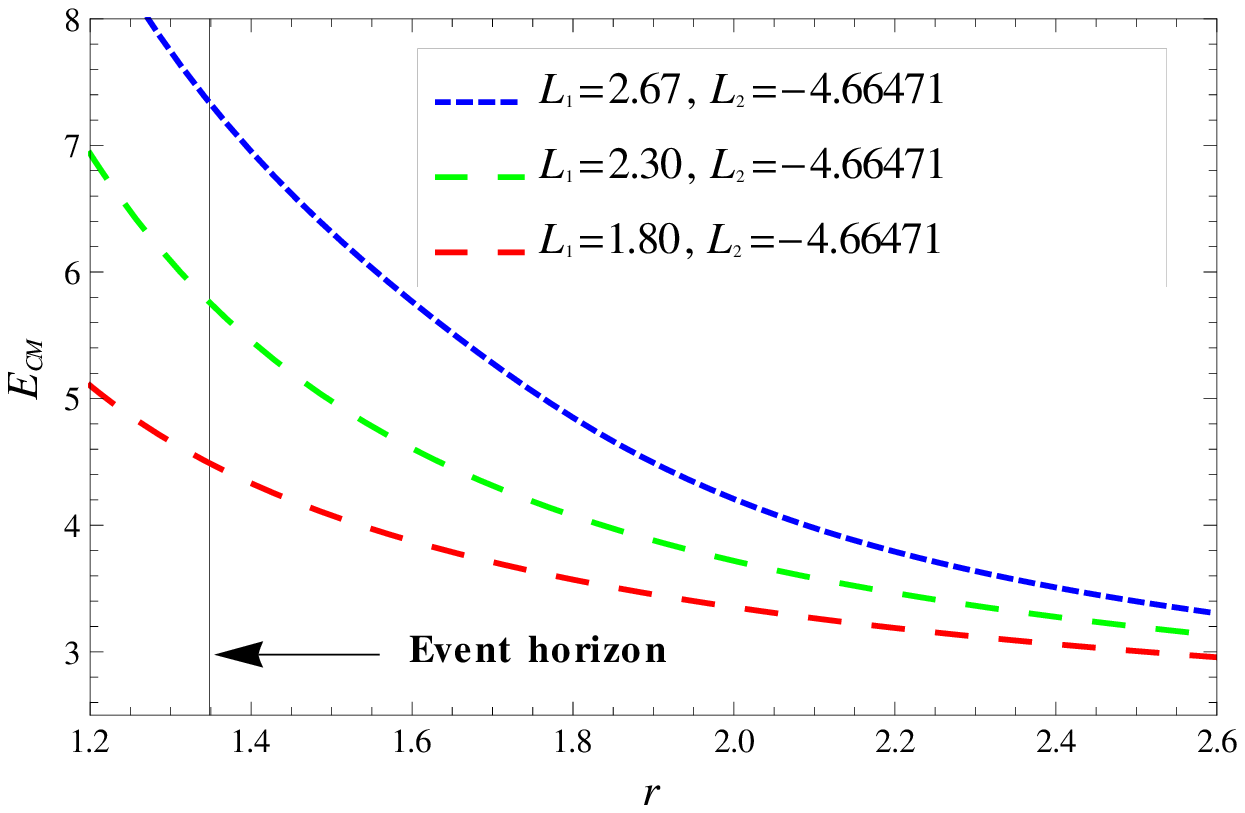}\hspace{-0.2cm}
&\includegraphics[scale=0.62]{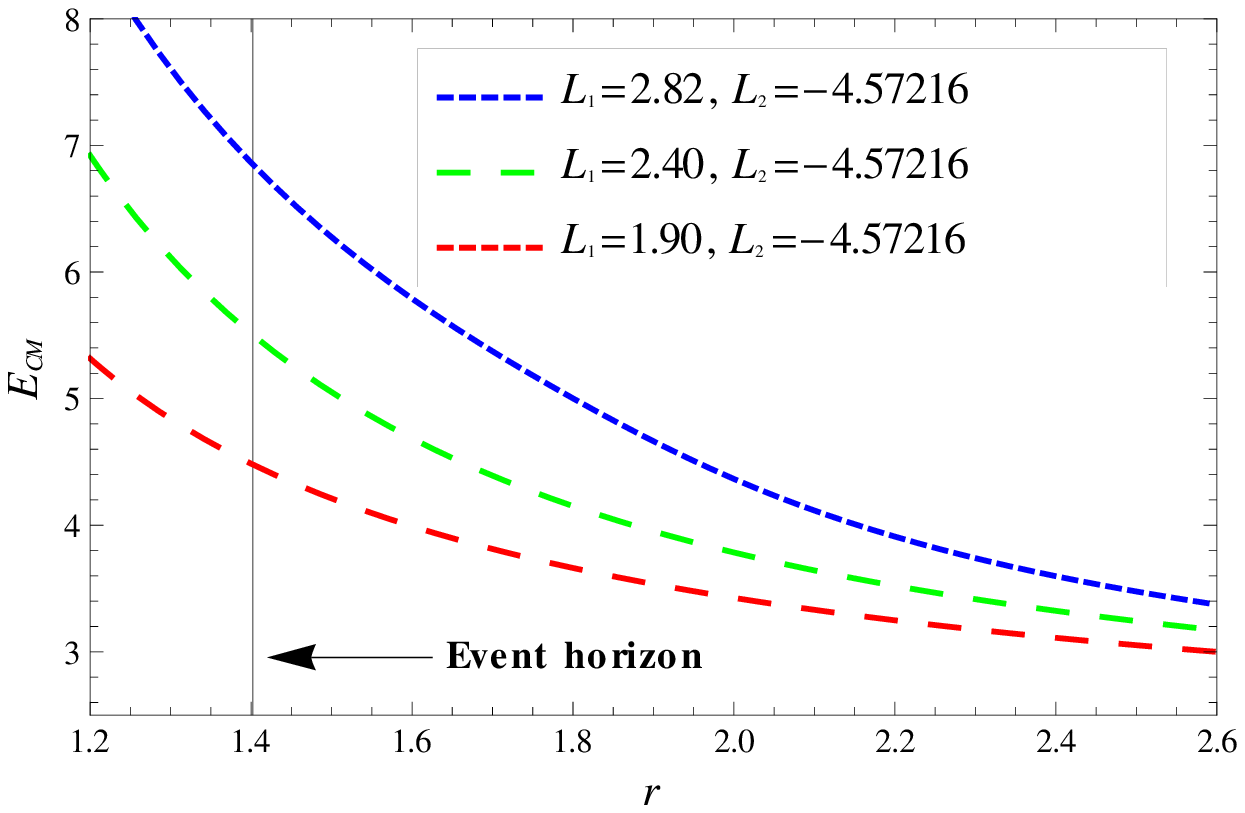}
\end{tabular}
 \caption{Plot showing the behavior of $E_{CM}$ vs $r$ for non-extremal black hole. (Top) For $a=0.95$, $g=0$ (left). For $a=0.9$, $g=0.2$ (right). (Bottom) For $a=0.8$, $g=0.3$ (left). For $a=0.7$, $g=0.4$ (right).}\label{fig11}
\end{figure*}

\section{Near horizon collision  in the rotating Bardeen regular black hole}
\label{cme}
Recently, Banãdos, Silk, and West (BSW) \cite{Banados:2009pr} analyzed the possibility that Kerr black hole can act as particle accelerator by studying collision of two particles near the event horizon of the Kerr black hole and found
that the center-of-mass energy ($E_{CM}$) of the colliding particles in the equatorial plane can be arbitrarily high in the limiting case of extremal black hole. Thus the extremal Kerr black hole can act as a Planck energy scale particle
accelerator.   This gave opening   to explore ultrahigh energy collisions and astrophysical phenomena, such as
gamma ray bursts and active galactic nuclei. Hence, the BSW mechanism received a significant attention  to study the collision of two particles near a rotating black hole \cite{Berti:2009bk,Banados:2010kn,Jacobson:2009zg,thorne,Zaslavskii:2010aw,Wei:2010gq,Hussain:2012zza,
Ghosh:2014mea,Amir:2015pja} (see also \cite{Harada:2014vka}, for  a review).
In this section, we want to study the properties of the $E_{CM}$ as $r$ tends to the event horizon $r^{+}_{H}$ in the case of the rotating Bardeen regular black hole.
Let us find the $E_{CM}$ for two colliding particles with the same rest mass $m_{1}= m_{2}=m_{0}$ are coming from rest at infinity. The collision energy in the centre-of-mass frame is defined as
\begin{equation}\label{eqlm}
E_{CM} = m_{0} \sqrt{2} \sqrt{1-g_{\mu \nu} u^{\mu}_{(1)} u^{\nu}_{(2)}},
\end{equation}
where $u^{\mu}_{(1)}$ and $u^{\nu}_{(2)}$ are the four-velocities of colliding particles. By using Eq.~(\ref{eqlm}) and substituting the values of four-velocities we can get the $E_{CM}$ for the black hole (\ref{bar}) as
\begin{eqnarray}\label{ecm}
\frac{E_{CM}^2}{2 m_{0}^2 } &=& \frac{1}{r(r^2-2mr+a^2)} \Big[2a^2 (m+r)
\nonumber \\
&-& 2am(L_{1}+L_{2})
\nonumber \\
&-&L_{1} L_{2}(-2m+r) +2(-m+r)r^2
\nonumber \\
&-&\sqrt{2m(a-L_{1})^2 - L_{1}^2r + 2mr^2}
\nonumber \\ &&
\sqrt{2m(a-L_{2})^2 - L_{2}^2r + 2mr^2}\Big].
\end{eqnarray}
Here $m=(M r^3)/(r^2+g^2)^{3/2}$. Thus the above equation is $g$ dependent, and when $g=0$, it will look exactly same with $m$ replaced by $M$, which is also exactly same obtained for the Kerr black hole \cite{Banados:2009pr}. Obviously as $r \rightarrow r_{H}^{E}$, Eq.~(\ref{ecm}) has the indeterminate form when we choose numerical values of $r_{H}^{E}$, $a$, $M$, and $g$. We apply l'Hospital's rule twice, then the value of the $E_{CM}$, as $r\rightarrow r_{H}^{E}$, becomes
\begin{eqnarray}\label{ecm1}
\frac{E_{CM}^2}{2 m_{0}^2 } (r\rightarrow r_{H}^{E}) &=& 8.08-0.51L_{1}L_{2}+0.48(L_{1}+L_{2})
\nonumber \\
&-& \frac{A_{1}(L_{2}-L_{c})}{2(L_{1}-L_{c})} -\frac{A_{2}(L_{1}-L_{c})}{2(L_{2}-L_{c})}
\nonumber \\
&-&\frac{B_{1}B_{2}}{3(L_{1}-L_{c})(L_{2}-L_{c})}+\frac{B_{1}^2(L_{2}-L_{c})}{6(L_{1}-L_{c})^3}
\nonumber \\
&+&\frac{B_{2}^2(L_{1}-L_{c})}{6(L_{2}-L_{c})^3},
\end{eqnarray}
with fixed $a=a_{E}=0.942439535325$, $r=r_{H}^{E}=1.04769$, $M=1$, and $g=0.2$.
The constants $A_{i}$ and $B_{i}$ corresponds to the $A_{i}=3.56+0.91L_{i}-0.51L_{i}^2$ and $B_{i}=4.35-0.35L_{i}-0.81L_{i}^2$ ($i=1,2$).
Eq.~(\ref{ecm1}) suggest that the unlimited $E_{CM}$ is possible if one the particle has critical angular momentum $L_c$. Therefore, $E_{CM}$ is divergent at the horizon $r=r_{E}$, if one of particles satisfies
the critical condition $E -  \Omega_{H} L_{c} = 0$.  The restriction on the $L_c$  is shown in the Table~(\ref{table3}). Numerical value of critical angular momentum $L_{c}= E/\Omega_{H}=2.10714$, which is exactly same as $L_{1}$. Hence, we can say that the $E_{CM}$ is infinite for the extremal rotating Bardeen regular black hole. We plot in Fig.~\ref{fig9},  $E_{CM}$ vs $r$ for various values of the angular momentum.

\subsubsection{Near horizon collision in non-extremal rotating Bardeen regular black hole}
Lastly, we study properties of $E_{CM}$ the limit $r \rightarrow r_{H}^{+}$ of non-extremal rotating Bardeen regular black hole.  Again as $r \rightarrow r_{H}^{+}$, both numerator and denominator  of the Eq.~(\ref{ecm}) vanish.  Hence, by application of l'Hospital's rule, we find that the $E_{CM}$, for near horizon for non-extremal rotating Bardeen black hole, reads
\begin{eqnarray}\label{ecm2}
\frac{E_{CM}^2}{2 m_{0}^2 } (r \rightarrow r_{H}^{+}) &=& \frac{1}{(L_{3}-L'_{c})(L_{4}-L'_{c})}\Big[12.14 \nonumber \\ &+&1.19(L_{3}^2+L_{4}^2)
\nonumber \\
&-&4.28(L_{3}+L_{4})- 0.86L_{3}L_{4}\Big].
\end{eqnarray}
In the above calculation we have fixed $a=0.9$, $r=r_{H}^{+}=1.31866$ and $g=0.2$.  The above Eq.~(\ref{ecm2}) is the formula for $E_{CM}$ of two colliding particle for non-extremal rotating Bardeen black hole.  The $E_{CM}$ will be infinite if either $L_{3}$ or $L_{4}$ equal to $L'_{c} = E/\Omega_{H}$, where $L_4 < L < L_3$  is the range for angular momentum with which a particle can reach horizon.
The critical value of angular momentum is calculated as $L'_{c}=2.83213$.  Hence, $L'_{c}$ is not in the acceptable range (cf. Table~\ref{table4}). Therefore, we can say that $E_{CM}$ is finite in the case of non-extremal black hole (cf. Figs.~\ref{fig10} and~\ref{fig11}).

\section{Conclusion}
\label{conclusion}
There has been many efforts to push as
much as possible Einstein gravity to its limit, trying to avoid the central singularity. Following some very early ideas  of Sakharov \cite{Sakharov:1966}, Gliner \cite{Gliner:1966} and Bardeen\cite{Bardeen:1968}, solutions possessing a
global structure alike to the one of black hole spacetimes, but in which the central singularity is absent, have been found. The first regular black hole solution having an event horizon was obtained by Bardeen \cite{Bardeen:1968}, which is a solution of Einstein equations in the presence of an electromagnetic field.  Recently, the rotating Bardeen regular black hole was also found \cite{Bambi:2013ufa}. Further, astrophysical black hole candidates are thought to be the Kerr black hole of general relativity, However, the actual nature of these objects is still to be verified.
In this paper, we have investigated in detail the horizons and ergosphere in the rotating Bardeen regular black hole and also analyzed the possibility that it can act as particle accelerator by studying the collision of two particles falling freely from rest at infinity.
The horizon structure of the rotating Bardeen regular black hole is complicated as compared to the Kerr black hole.
It has been observed that, for each $g$ and suitable choice of parameters, we can find critical value $a=a_{E}$, which corresponds to an extremal black hole with degenerate horizons, i.e., when  $a=a_{E}$, the two horizons coincides (cf. Figs.~\ref{fig1} and ~\ref{fig2}). Interestingly, the value $a_{E}$ is sensitive to the parameter $g$, $a_{E}$ decreases with increase in $g$. However, when $a<a_{E}$, we have a regular black hole with Cauchy and event horizon, and for $a>a_{E}$, no horizon exists. Further, the ergosphere area increases with increase in $a$. On the other hand, when the value of $g$ increases, the ergosphere area enlarges. It has been suggested that ergosphere can be used to extract energy from rotating black holes through the Penrose process \cite{Penrose}. Howevever, Penrose himself said
that the method is inefficient \cite{Penrose1}, although later \cite{Chandra} he showed that the theoretical efficiency could reach 20\% extra energy up to 60\%. In the case of rotating black hole the efficiency of the collisional process is not high, if a magnetic field effect is not involved rather than it is efficient in the case of Kerr naked singularity \cite{Stuchlik:2013yca, Patil:2014lea}. Hence, the parameter $g$ may play significant role in the energy extraction process from the rotating Bardeen regular black holes that is being investigated separately.

It has been shown by BSW \cite{Banados:2009pr} that the $E_{CM}$ of two colliding particle may occur at arbitrarily high energy for the case of extremal Kerr black holes. The BSW analysis when extended to the rotating Bardeen regular black hole, with some restriction on parameters, the arbitrary high $E_{CM}$ can be achieved when collision takes place near horizon of an extremal rotating Bardeen regular black hole with one of the colliding particle must have critical angular momentum. We have also calculated the range of angular momentum for which a particle particle reach the horizon  i.e., if the angular momentum lies in the range, the collision is possible near the horizon of the rotating Bardeen regular black holes.   The  calculations of $E_{CM}$, was performed in the case of the the rotating Bardeen regular black hole for various values of $g$, and we found that  they are infinite if one of the colliding particles has the critical angular momentum in the required range. On the other hand, the $E_{CM}$ has always finite upper limit for the non-extremal rotating Bardeen regular black hole. Thus, the BSW mechanism, for the rotating Bardeen regular black hole, depends both on the rotation parameter $a$ as well as on the deviation parameter $g$. For the non-extremal black hole, we have also seen the effect of $g$ on the $E_{CM}$ demonstrate a increase in the value of the $E_{CM}$ with an increase in the value of $g$.

According to the no-hair theorem, all astrophysical black holes are
expected to be like Kerr black holes, but the actual nature of this has still to be verified \cite{Bambi:2014nta}. The impact of the parameter $g$  on the the horizon structure, ergroregion and particle acceleration  presents a good theoretical opportunity to distinguish the Bardeen-Kerr regular black hole from the Kerr black and to test whether astrophysical black hole candidates are the black holes as predicted by Einstein's general relativity.

\begin{acknowledgements}
M.A. acknowledges the University Grant Commission, India, for financial support through the Maulana Azad National Fellowship For Minority Students scheme (Grant No.~F1-17.1/2012-13/MANF-2012-13-MUS-RAJ-8679).  S.G.G. would like to thank SERB-DST for Research Project Grant NO SB/S2/HEP-008/2014, also would like to thank IUCAA, pune for hospitality while part of this work was done.
\end{acknowledgements}



\begin{thebibliography}{99}

\bibitem{Reissner:1916}
H.~Reissner,
Ann.\ Phys.\ {\bf 355}, 9 (1916) .

\bibitem{Hawking:1970}
  S.~W.~Hawking and R.~Penrose,
  Proc.\ R.\ Soc.\ A {\bf 314}, 529 (1970).

\bibitem{Hawking:1973}
  S.~W.~Hawking and G.~F.~R.~Ellis,
  The Large Scale Structure of Space and Time
  (Cambridge University Press, Cambridge, 1973).

\bibitem{Wheeler:1964}
J. A. Wheeler,
 in Relativity, Groups, and Topology,
 edited by C. DeWitt and B. DeWitt
 (Gordon and Breach, New York, 1964), p. 315.

\bibitem{Sakharov:1966}
A. D. Sakharov,
 Sov.\ Phys.\ JETP, {\bf 22}, 241 (1966).

\bibitem{Gliner:1966}
E. B. Gliner,
 Sov.\ Phys.\ JETP, {\bf 22}, 378 (1966).

\bibitem{Bardeen:1968}
J.~Bardeen,
in {\it Proceedings of GR5} (Tiflis, U.S.S.R., 1968).

\bibitem{Borde:1994ai}
  A.~Borde,
  Phys.\ Rev.\ D {\bf 50}, 3692 (1994).

\bibitem{Borde:1996df}
  A.~Borde,
  Phys.\ Rev.\ D {\bf 55}, 7615 (1997).

\bibitem{AyonBeato:1998ub}
  E.~Ayon-Beato and A.~Garcia,
  Phys.\ Rev.\ Lett.\  {\bf 80}, 5056 (1998).

\bibitem{Bronnikov:2000vy}
  K.~A.~Bronnikov,
  Phys.\ Rev.\ D {\bf 63}, 044005 (2001).

\bibitem{Hayward:2005gi}
  S.~A.~Hayward,
  Phys.\ Rev.\ Lett.\  {\bf 96}, 031103 (2006).

\bibitem{Zaslavskii:2009kp}
  O.~B.~Zaslavskii,
  Phys.\ Rev.\ D {\bf 80}, 064034 (2009).

\bibitem{Lemos:2011dq}
  J.~P.~S.~Lemos and V.~T.~Zanchin,
  Phys.\ Rev.\ D {\bf 83}, 124005 (2011).

\bibitem{Stuchlik:2014qja} 
  Z.~Stuchlík and J.~Schee,
  Int.\ J.\ Mod.\ Phys.\ D {\bf 24}, 1550020 (2014).

\bibitem{Schee:2015nua} 
  J.~Schee and Z.~Stuchlik,
  JCAP {\bf 1506}, 048 (2015).
    
\bibitem{Bambi:2013ufa}
  C.~Bambi and L.~Modesto,
  Phys.\ Lett.\ B {\bf 721}, 329 (2013).

\bibitem{Bambi:2014nta}
  C.~Bambi,
  Phys.\ Lett.\ B {\bf 730}, 59 (2014).

\bibitem{Ghosh:2014hea}
  S.~G.~Ghosh and S.~D.~Maharaj,
  Eur.\ Phys.\ J.\ C {\bf 75}, no. 1, 7 (2015)

\bibitem{Amir:2015pja}
  M.~Amir and S.~G.~Ghosh,
  arXiv:1503.08553.

\bibitem{Ghosh:2014pba}
  S.~G.~Ghosh,
  arXiv:1408.5668.

\bibitem{Toshmatov:2014nya}
  B.~Toshmatov, B.~Ahmedov, A.~Abdujabbarov and Z.~Stuchlik,
  Phys.\ Rev.\ D {\bf 89}, 104017 (2014).

\bibitem{Gou:2011}
L.~Gou,
Astrophys.\ J.\ {\bf 742}, 85 (2011).

\bibitem{Gou:2013dna}
  L.~Gou,
  arXiv:1308.4760.

\bibitem{Li:2013jra}
  Z.~Li and C.~Bambi,
  JCAP {\bf 1401}, 041 (2014).

\bibitem{Newman:1965tw}
  E.~T.~Newman and A.~I.~Janis,
  J.\ Math.\ Phys.\  {\bf 6},  915 (1965).

\bibitem{thorne}
K. S. Thorne, Astrophys. J. {\bf 191}, 507 (1974).

\bibitem{Penrose} R. Penrose and R. M. Floyd, Nature Physical Science 229, 177 (1971).

\bibitem{Chandra} S. Chandrasekhar, The Mathematical Theory of Black Holes
(Oxford University Press, Oxford, 1983).

\bibitem{Banados:2009pr}
M.~ Ban\~{a}dos, J.~Silk and S.~M.~West,
Phys.\ Rev.\ Lett.\  {\bf 103}, 111102 (2009).

\bibitem{Jacobson:2009zg}
T.~Jacobson and T.~P.~Sotiriou,
Phys.\ Rev.\ Lett.\  {\bf 104}, 021101 (2010).

\bibitem{Berti:2009bk}
  E.~Berti, V.~Cardoso, L.~Gualtieri, F.~Pretorius and U.~Sperhake,
  Phys.\ Rev.\ Lett.\  {\bf 103}, 239001 (2009).

\bibitem{Banados:2010kn}
  M.~Ban\~{a}dos, B.~Hassanain, J.~Silk and S.~M.~West,
  Phys.\ Rev.\ D {\bf 83}, 023004 (2011).

\bibitem{Wei:2010gq}
  S.~W.~Wei, Y.~X.~Liu, H.~T.~Li and F.~W.~Chen,
  J. High Energy Phys. {\bf 12} (2010) 066.

\bibitem{Hussain:2012zza}
  I.~Hussain,
  Mod.\ Phys.\ Lett.\ A {\bf 27}, 1250017 (2012).

\bibitem{Ghosh:2014mea}
  S.~G.~Ghosh, P.~Sheoran and M.~Amir,
  Phys.\ Rev.\ D {\bf 90}, no. 10, 103006 (2014).

\bibitem{Zaslavskii:2010aw}
  O.~B.~Zaslavskii,
  JETP Lett.\  {\bf 92}, 571 (2010)
  [Pisma Zh.\ Eksp.\ Teor.\ Fiz.\  {\bf 92}, 635 (2010)].

\bibitem{Harada:2014vka}
  T.~Harada and M.~Kimura,
  Classical\ Quantum\ Gravity\  {\bf 31}, 243001 (2014).

\bibitem{Penrose1} R. Penrose, Riv. Nuovo Cimento Numero Speciale 1, 252
(1969) [Gen. Relativ. Gravit. 34, 1141 (2002)].

\bibitem{Patil:2014lea} 
  M.~Patil and P.~S.~Joshi,
  Pramana {\bf 84}, 491 (2015).
  
\bibitem{Stuchlik:2013yca} 
  Z.~Stuchlik and J.~Schee,
  Class.\ Quant.\ Grav.\  {\bf 30}, 075012 (2013).  
  

\end{thebibliography}
\end{document}